\documentclass{aa}

\usepackage{algorithm}
\usepackage[end,]{algpseudocode}
\usepackage{amssymb}
\usepackage{caption}
\usepackage{graphicx}
\usepackage{hyperref}
\usepackage{mathtools}
\usepackage{natbib}
\usepackage{setspace}
\usepackage{siunitx}
\usepackage{subcaption}
\usepackage{txfonts}
\usepackage{xcolor}

\algnewcommand{\LineComment}[1]{\Statex \(\triangleright\) #1}

\usepackage{xparse}
\makeatletter
\NewDocumentCommand{\LeftComment}{s m}{%
  \Statex \IfBooleanF{#1}{\hspace*{\ALG@thistlm}}\(\triangleright\) #2}
\makeatother
  
\begin{document}

   \title{Simulating images of radio galaxies with diffusion models}


   \author{
        T. Vičánek Martínez \inst{1}
        \and N. Baron Perez \inst{1} \and M. Brüggen \inst{1}
    }
    
   \institute{
        Hamburger Sternwarte, Universität Hamburg, Gojenbergsweg 112, 21029 Hamburg, Germany
   }

   \date{}

  \abstract
   {With increasing amounts of data produced by astronomical surveys, automated analysis methods have become crucial. Synthetic data are required for developing and testing such methods. Current classical approaches to simulations often suffer from insufficient detail or inaccurate representation of source type occurrences. Deep generative modeling has emerged as a novel way of synthesizing realistic image data to overcome those deficiencies.}
   {We implemented a deep generative model trained on observations to generate realistic radio galaxy images with full control over the flux and source morphology.}
   {We used a diffusion model, trained with continuous time steps to reduce sampling time without quality impairments. The two models were trained on two different datasets, respectively. One set was a selection of images obtained from the second data release of the LOFAR Two-Metre Sky Survey (LoTSS). The model was conditioned on peak flux values to preserve signal intensity information after re-scaling image pixel values. The other, smaller set was obtained from the Very Large Array (VLA) survey of Faint Images of the Radio Sky at Twenty-Centimeters (FIRST). In that set, every image was provided with a morphological class label the corresponding model was conditioned on. Conditioned sampling is realized with classifier-free diffusion guidance. We evaluated the quality of generated images by comparing the distributions of different quantities over the real and generated data, including results from the standard source-finding algorithms. The class conditioning was evaluated by training a classifier and comparing its performance on both  real and generated data.}
   {We have been able to generate realistic images of high quality using 25 sampling steps, which is unprecedented in the field of radio astronomy. The generated images are visually indistinguishable from the training data and the distributions of different image metrics were successfully replicated. The classifier is shown to perform equally well for real and generated images, indicating strong sampling control over morphological source properties.}
   {}


   \maketitle

\section{Introduction}
Sky surveys have become essential to modern astronomy with respect to improving our understanding of the universe, as they significantly expand the amount of recorded signals available to scientific exploration. Their growing range and sensitivity have brought on an exponential increase in data volumes \citep{Smith&Geach23}. For instance,   LOFAR Two-Metre Sky Survey will gather a total amount of about $\SI{50}{PB}$ of data \citep{Shimwell+2017}, while Square Kilometre Array (SKA) is expected to yield approximately $\SI{4.6}{EB}$ \citep{Zhang&Zhao15}. As conventional analysis often relies on human supervision and expert knowledge, many commonly used techniques become intractable for larger amounts of data. This entails an inevitable demand for automated methods to extract useful information at scale and thereby facilitate scientific discoveries. In the case of radio observations, this challenge is manifested in tasks such as source detection, image segmentation, and morphological classifications. Those are frequently obstructed by high levels of background noise and nearby artifacts arising from the imaging process, or complicated by the fact that many sources show multiple non-contiguous islands of emission that are often ambiguous and not easy to associate with each other.\\

In this context, the use of synthetic data that are applied in several different ways has become particularly valuable. For instance, tools for automatic source extraction are tested on simulated image data \citep{HydraII}, which makes it possible to evaluate the results against properties introduced in the simulation. Furthermore, simulated observations have also been employed in developing deep learning-based methods for automatic calibration \citep{SmartCalibration} and system health inspection \citep{Mesarcik+2020} of modern radio telescopes. Simulated data are also commonly employed in testing methods for radio image reconstruction, as, for instance, in \cite{LOFAR_sparse}. Another use case for synthetic radio images explored by \cite{Rustige+2023} is the augmentation of available training data required for deep learning methods such as morphological classification, where data sets are often insufficiently sized or highly imbalanced.\\
In the absence of an analytical model for radio galaxy formation and evolution, generating synthetic images that are representative of real observations is not a straightforward task. Simulations often rely on the assumption of an ideal background and extended sources are typically modeled with geometrical shapes. For instance, \cite{Wilman+2008} simulate flux distributions of extended sources as combinations of point source cores or hot-spots, and elliptical lobes. Similarly, \cite{HydraI} represent extended sources as composites of 2D Gaussian components. These representations are insufficient to capture the morphological complexity of real extended sources. Furthermore, in addition to properties of the signal, an accurate portrayal of astronomical observations also requires detailed knowledge of telescope properties, observation conditions, and data processing. A different approach, as employed in the first SKA Data Challenge \citep{SKA_Ch1_results}, consists of using real high-resolution images enhanced with random transformations, such as rotations and scaling that provide further variability. While resulting in more accurate representations, this approach lacks diversity in the generated images and can lead to an under-representation of different morphological sub-classes, inherited from the corresponding distribution present in the applied image library.\\

A promising way to overcome these limitations is through the use of deep generative models. As the field of image synthesis with methods of machine learning has experienced tremendous progress in recent years in the domain of natural images, those techniques have also found their way into the field of radio astronomy. Generative models are deep learning models that learn to represent the underlying probability distribution of the data sets they are trained on. Once that training is completed, they can be used to synthesize new data points that follow the learned distribution. This facilitates the generation of realistic artificial images containing complex structures, without the need for explicit modeling. Imaging properties of the telescope are implicitly learned from the training images. In addition, the possibility of conditioned sampling allows us to control of different image parameters, such as the morphological class of depicted sources, thereby circumventing the potential issue of under-representation.\\
A particular type of deep generative models that has experienced a recent rise in popularity across different domains is called score-based models or, more commonly, diffusion models (DMs). These models make use of neural networks that at first are trained to remove Gaussian noise, added during the training process, from images. Once trained, this neural network is then employed in an iterative fashion to morph a seed array of pure Gaussian noise into an image that is equivalent to a sample drawn from the distribution underlying the training dataset. In the domain of natural imaging, this class of generative models has emerged as state-of-the-art in terms of sample quality, outperforming other common approaches \citep{Dhariwal+21} such as variational autoencoders (VAEs) or generative adversarial networks (GANs).\\

While different approaches to generative modeling have already been employed in the context of astronomical imaging (e.g. \cite{Bastien+21}), DMs have been explored more recently across a range of contexts. \cite{Smith+2022} implemented a DM trained on optical galaxy observations and demonstrated the similarity of generated images to the real data. In addition, they demonstrated the potential use of DMs for image in-painting, using their model to remove simulated satellite trails from images and reliably reconstruct parts of the image. They further show the use of domain transfer, where the model is utilized to make cartoon images look like real observations with similar features, giving an idea of the degree of control that can be exerted over the sampling process of a DM.\\
\cite{Sortino+2023} implemented a latent DM, namely, a DM that acts on the latent space of an autoencoder to generate synthetic images of radio sources. The model is conditioned on segmentation maps and reference images, allowing the user to control the shape of the source and the background properties of the image, which has been shown to  work reliably. While this facilitates radio image synthesis with explicit image properties, this approach requires a background reference image together with a corresponding semantic map for every generated sample, which is not easy to obtain at scale.\\

\cite{Zhao+2023} used simulated images of \SI{21}{cm} temperature mapping to train both a DM and a GAN, conditioned on cosmological parameters that influence the result of the simulation. They quantitatively demonstrated the superiority of DMs over GANs in terms of image quality, as well as the recovery of the properties determined by the conditioning parameters.\\
Apart from image generation, \cite{Wang+2023} and \cite{Drozdova+2024} demonstrated the successful use of DMs for the reconstruction of radio interferometric images. \cite{WaldmannRocchetto2023} used a DM to remove noise and artifacts from optical observations of satellites and space debris. In a recent paper, \cite{Reddy+24} used DMs for implementing a super-resolution model for gravitational lensing data, which allows for both the lensing system and the background source to be understood in more detail. Their studies show that the DM outperforms existing techniques. In this work, we introduce a DM that can generate realistic radio galaxy images over a wide dynamic range. In an approach novel to radio astronomy, we employed a recent framework referred to as continuous-time DMs, which allows for a flexible sampling procedure with a reduced number of steps. 

Whenever we mention radio images in this paper, we mean cleaned and deconvolved images, since these are the images that we train on. The generation of synthetic radio data that includes the dirty images (which are the initial images of the sky produced directly by Fourier transforms of the raw data) is done elsewhere \cite[e.g.]{2023A&A...677A.167G}. The dirty images obviously depend on the telescope and the parameters of each individual observation, such as the $uv$-coverage. Hence we also do not deal with the imperfections in the dirty image that are often dominated by sidelobes and other artifacts. This is discussed further in Sect.~\ref{sec:discussion}.

We demonstrate control over the morphological properties of the samples by training a DM conditioned on morphological class labels. In Sect. \ref{sec:diffusion-models} we introduce the theory behind DMs. Section \ref{sec:training-data} describes the gathering and pre-processing of the image data used to train our models. In Sect. \ref{sec:model-and-training} we describe the neural network architecture, as well as the training and sampling procedures. Finally, our metrics used to evaluate the model performance are defined in Sect. \ref{sec:evaluation-metrics}. The results are presented in Sect. \ref{sec:results} and discussed in Sect. \ref{sec:discussion}.\\
   
\section{Diffusion models} 
\label{sec:diffusion-models}

A DM transforms images of pure Gaussian noise into images that look as if they had been drawn from the training data set. Implicitly, this constitutes a mapping between two probability distributions in image space: a normal Gaussian distribution on one hand and a distribution that models the training dataset, on the other. The underlying theory is derived by first considering the opposite direction, a gradual diffusion of a training image into pure noise, referred to as the 'forward process' that is trivial to accomplish in practice. With the correct description, this process becomes time-reversible and the desired backward process can be approximated at discretized time steps using a denoising neural network.\\
The concept was first introduced in \citet{Sohl-Dickenstein+15} and further developed in \citet{Ho+20} and subsequent studies. In those works, the model is trained on fixed discrete time steps of the diffusion process, which are also used for sampling. Later, \cite{Song+20} found a framework that unified different approaches to DM training, and \cite{Karras+22} proceeded to show the benefits of training the model with random continuous time values sampled from a distribution. In this work, we implement such a "continuous-time" DM and this section presents the underlying theory.

\subsection{Forward process and probability flow ODE}
For a vector $\mathbf{x} \in \mathbb{R}^{d\times d}$ in the space of all images with $d\! \times\! d$ pixels, the forward process of gradually diffusing an image with Gaussian noise corresponds to a probabilistic trajectory through that space. \cite{Song+20} introduce a modeling of the forward process $\mathbf{x}(t)$ as a function of a continuous time variable $t \in [0, T]$ expressed by a stochastic differential equation (SDE)

\begin{equation}
    \mathrm{d} \mathbf{x}=\mathbf{f}(\mathbf{x}, t) \mathrm{d} t+g(t) \mathrm{d} \mathbf{w}, \label{eq:SDE}
\end{equation}
where $\mathbf{f}$ and $g$ are called the drift and diffusion coefficients, respectively. Those two functions are free parameters that can be chosen to construct a process with desired properties that serve different practical purposes. Also, $\mathbf{w}$ is the standard Wiener process, also referred to as Brownian motion, which describes continuous-time independent Gaussian increments of zero mean and unit variance. If $\mathbf{x}(0)$ is sampled from an initial distribution $p_0(\mathbf{x})$, which in this context is the distribution underlying the training data, then Eq. \eqref{eq:SDE} defines a marginal probability density $p_t(\mathbf{x})$ for $\mathbf{x}(t)$ at every point in time given by:

\begin{equation}
    p_t(\mathbf{x}) \coloneqq \int p_0(\mathbf{x}') \cdot \mathcal{N}\left(\mathbf{x}\,|\,\mathbf{x}', \sigma(t)^2\mathbb{I}\right) \;\mathrm{d}\mathbf{x}', \label{eq:marginalProb}
\end{equation}
where $\mathcal{N}\left(\mathbf{x}\,|\,\mathbf{x}', \sigma(t)^2\mathbb{I}\right)$ is the Gaussian normal distribution centered at $\mathbf{x}'$ with a standard deviation of $\sigma(t)$, the value of which depends on the choices of $\mathbf{f}$ and $g$. In the context of DMs, $\sigma$ is referred to as the noise level, and the function $\sigma(t)$ is called the noise schedule. We note that $p_t(\mathbf{x})$ is always dependent on the initial distribution $p_0(\mathbf{x})$; however, for simplicity we omit this relation in the corresponding expression.\\
In the same work, the authors also derive an equivalent ordinary differential equation (ODE) for every SDE as expressed in Eq. \eqref{eq:SDE}, given by

\begin{equation}
    \mathrm{d} \mathbf{x}=\left[\mathbf{f}(\mathbf{x}, t)-\frac{1}{2} g(t)^2 \nabla_{\mathbf{x}} \log p_t(\mathbf{x})\right] \mathrm{d} t, \label{eq:PF-ODE}
\end{equation}
 which is named probability flow ODE (PF-ODE). The term $\nabla_{\mathbf{x}} \log p_t(\mathbf{x})$ is referred to as the score function, and can be thought of as a vector field pointing in the direction of increasing probability 
 density of $p_t(\mathbf{x})$. Equation \eqref{eq:PF-ODE} is equivalent to Eq. \eqref{eq:SDE} in the sense that the solving trajectories share the same marginal probabilities $p_t(\mathbf{x})$ at every point in time. However, the advantage of using the PF-ODE is that the process described therein is not probabilistic but deterministic and, therefore, time-reversible. Consequently, if the score function is known, the process can be described in both directions of time by solving Eq. \eqref{eq:PF-ODE}. For a large enough $T$, the vector $\mathbf{x}(T)$ is practically indistinguishable from pure Gaussian noise, in other terms, $p_T(\mathbf{x}) \approx \mathcal{N}(0, \sigma(T)^2\mathbb{I})$. Thereby, Eq. \eqref{eq:PF-ODE} defines a deterministic mapping between a Gaussian noise vector $\mathbf{x}_T$ and an initial image $\mathbf{x}_0 \sim p_0(\mathbf{x})$. As we will discuss in the following section, it is possible to train a neural network to estimate the score function, and images can then be sampled from $p_0(\mathbf{x})$ by starting with random Gaussian noise and numerically solving the PF-ODE.\\
 In the scope of a more general description, \cite{Karras+22} give a different formulation of the PF-ODE. Choosing $\mathbf{f} \equiv 0$ and re-writing $g$ as an expression of $\sigma(t)$, Eq. \eqref{eq:PF-ODE} becomes
\begin{equation}
     \mathrm{d} \mathbf{x}=-\dot{\sigma}(t) \, \sigma(t) \, \nabla_{\mathbf{x}} \log p_t(\mathbf{x}) \;\mathrm{d} t, \label{eq:PF-ODE_Karras}
\end{equation}
enabling us to directly construct the process according to the desired noise schedule, which is chosen to optimize the intended goal, typically high quality of sampled images.

\subsection{Reverse process and denoising score matching}

\cite{Karras+22} proceed to show that the score function can be expressed through an ideal denoiser function. If $D(\mathbf{x};\sigma)$ is a function that, for every $\sigma(t)$ with $t \in [0, T]$ separately, minimizes the expected $L_2$ denoising loss over the initial distribution $p_0(\mathbf{x})$ expressed as 

\begin{equation}
\mathbb{E}L_2 = \mathbb{E}_{\mathbf{x} \sim p_0} \mathbb{E}_{\mathbf{n} \sim \mathcal{N}\left(\mathbf{0}, \sigma^2 \mathcal{I}\right)}\|D(\mathbf{x}+\mathbf{n} ; \sigma)-\mathbf{x}\|_2^2, \label{eq:denoiserLoss}
\end{equation}
then the score function can be written as

\begin{equation}
    \nabla_{\mathbf{x}} \log p_t(\mathbf{x}) = \frac{D(\mathbf{x}; \sigma(t)) - \mathbf{x}}{\sigma(t)^2}. \label{eq:scoreMatching}
\end{equation}
This denoiser function can be approximated by a neural network $D_\theta$, trained to minimize Eq. \eqref{eq:denoiserLoss} over a dataset of real images. In this way, the network implicitly models a probability distribution for the image data it was trained on. This distribution can then be sampled from by solving the equation obtained substituting Eq. \eqref{eq:scoreMatching} into Eq. \eqref{eq:PF-ODE_Karras}, which results in
\begin{equation}
    \mathrm{d} \mathbf{x}=-\frac{\dot{\sigma}(t)}{\sigma(t)} \, \left( D(\mathbf{x}; \sigma(t)) - \mathbf{x} \right) \;\mathrm{d} t. \label{eq:scoreMatchedODE}
\end{equation}
In practice, Eq. \eqref{eq:scoreMatchedODE} is solved through discretization, where the denoiser function is evaluated using $D_\theta$. This can in principle be done with any ODE solver, our implementation is described in Sect. \ref{sec:sub:Sampling}.

\subsection{Guided diffusion} 
\label{sec:sub:guided_diffusion}

In addition to $\mathbf{x}$ and $\sigma$, the neural network $D_\theta$ can be designed as a function of additional parameters $c$ that are passed as input, such as class labels describing the morphological class of the imaged galaxy for every training sample. This technique is referred to as conditioning, since the model then describes conditional probabilities $p_t(\mathbf{x}\,|\,c)$ that depend on the added parameters. In order to achieve high quality for conditional sampling, \cite{Ho&Salimans22} introduce the concept of classifier-free diffusion guidance, which builds on the idea of classifier guidance described in \cite{Dhariwal+21}. For classifier-free guidance, the network is simultaneously trained with and without conditioning by randomly dropping out the additional input $c$ during training and replacing it with a null token $\emptyset$ that has no effect on the network. Once trained, the denoiser is evaluated as a linear combination $\widetilde{D}_\theta$ of both the conditioned and unconditioned model. This is expressed by
\begin{equation}
    \widetilde{D}_\theta\left(\mathbf{x}; \sigma \,|\, c\right) = (1 + \omega) \cdot {D}_\theta\left(\mathbf{x}; \sigma \,|\, c\right) + \omega \cdot {D}_\theta\left(\mathbf{x}; \sigma \,|\, \emptyset\right), \label{eq:guidance}
\end{equation}
where $\omega$ is called guidance strength and is a parameter that can be chosen to optimize the desired output. \cite{Ho&Salimans22} show that the choice of $\omega$ presents a trade-off between sample variety for lower values and sample fidelity for higher values. 

\section{Training data} 
\label{sec:training-data}

This work is carried out on two different sets of radio galaxy images. One is an extensive dataset with radio galaxy images gathered from observations made with the LOFAR telescope. The other is the FIRST dataset from \cite{Griese+23}, a smaller set with images obtained from observations with the Very Large Array (VLA) telescope that was provided with labels which categorize every image into one of four morphological classes. This section describes how those datasets are assembled, processed and prepared for training.

\subsection{LOFAR dataset} \label{sec:sub:LOFAR_Data}

\begin{figure}
    \centering
    \resizebox{\hsize}{!}{\includegraphics{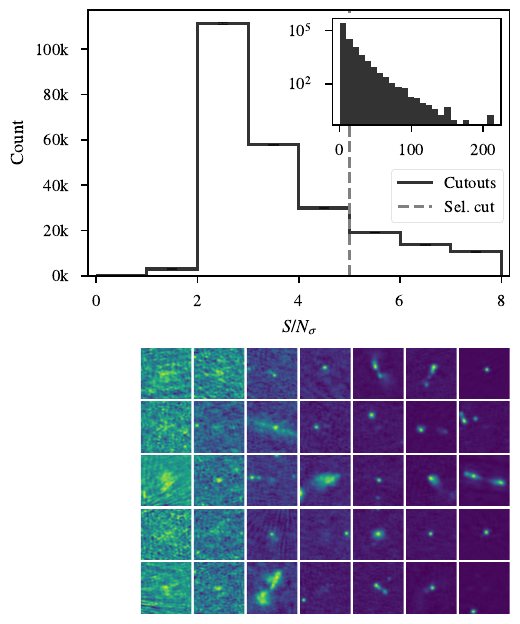}}
    \caption{Distribution of $\mathit{S/N}_\sigma$ over the entire set of cut-outs extracted from the LOFAR continuum map. The top panel shows the distribution over a range relevant to the selection cut threshold, while the inset figure in the top right shows the entire distribution. The bottom grid shows random examples out of the different histogram bins, stacked vertically and aligned horizontally with the corresponding bin.}
    \label{fig:sigma-S/N_examples}
\end{figure}

\begin{figure}
    \centering
    \resizebox{\hsize}{!}{\includegraphics{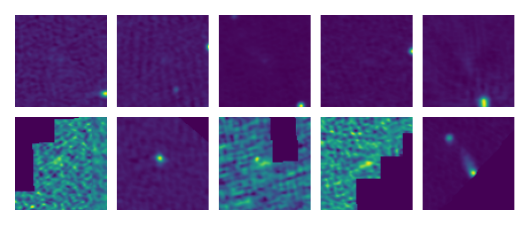}}
    \caption{Examples of cut-outs excluded from the training set due to bright edge pixels (top) or incomplete coverage (bottom).}
    \label{fig:excluded_examples}
\end{figure}

\begin{figure}
    \centering
    \resizebox{\hsize}{!}{\includegraphics{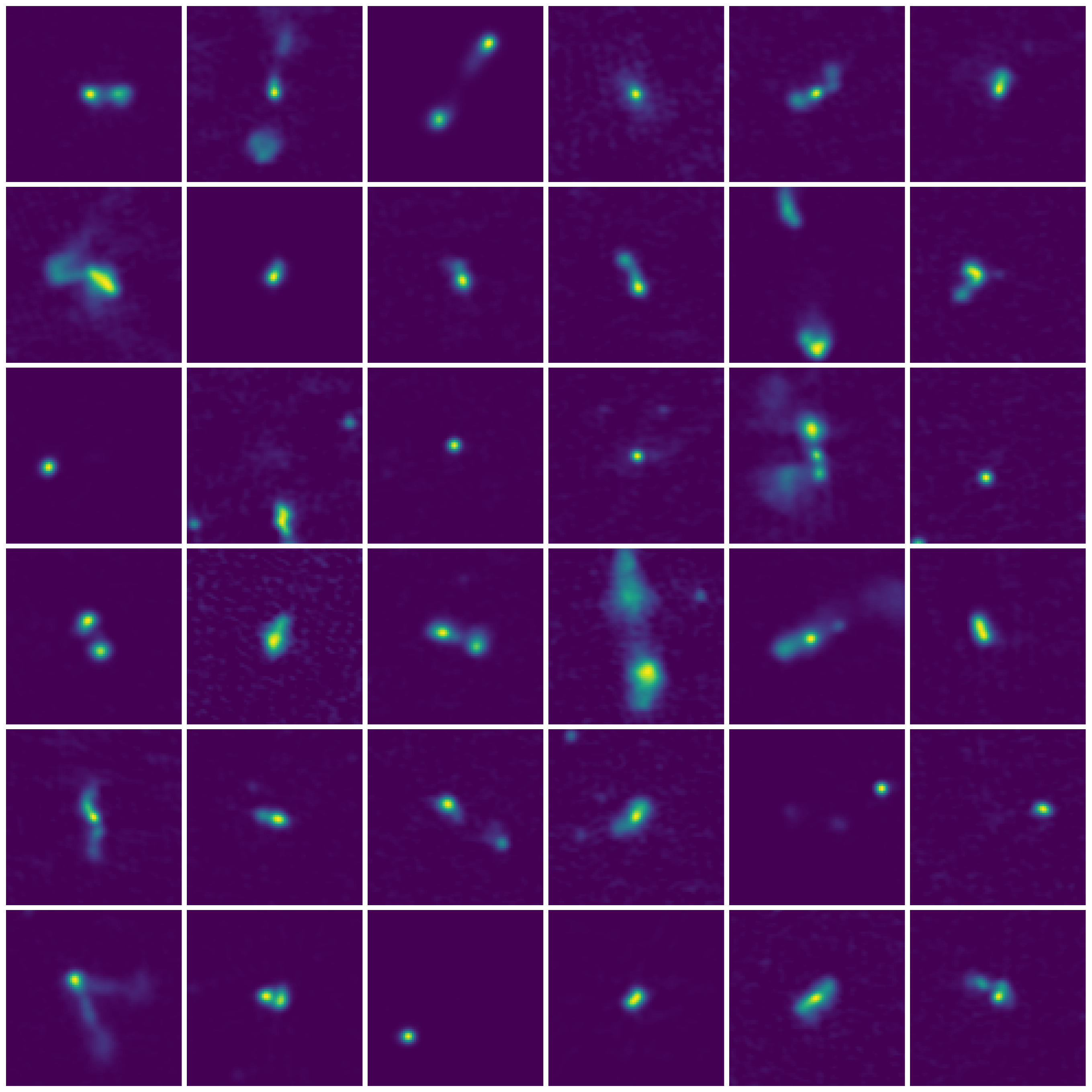}}
    \caption{Random examples of images included in the final LOFAR dataset.}
    \label{fig:selection_examples}
\end{figure}

The unlabeled LOFAR dataset used to train our LOFAR model is based on the second data release of the LOFAR Two-metre Sky Survey (LoTSS-DR2), a survey covering 27\% of the northern sky in a band of $\SI{120}{-}\SI{168}{MHz}$ with a central frequency of $\SI{144}{MHz}$, details are given in \cite{Shimwell+22}. We retrieve the publicly available mosaics of the Stokes I continuum maps as the base to draw images of single sources from, alongside with the corresponding optical cross-match catalog described in \cite{Hardcastle+23} that lists known sources and contains information on different characteristics thereof.\\
Cut-outs from the maps are taken around all resolved sources listed in the catalog, which is a total of \num{314942}. The size of the cut-outs is $\ang{;;120}$, which at the given resolution of $\SI{1.5}{\arcsecond\per{px}}$ corresponds to images of $80\! \times\! 80$ pixels in size. The cut-outs are centered around the coordinates of the optical counterparts of each radio source; however, if such a source is not present in the catalog, they are centered around the corresponding position of the radio source itself. We removed cut-outs that contain NaN or blank image values, referred to as "broken images" in this work.\\
To filter out images with a high degree of background noise or artifacts, which cause the source not to be distinctly visible on the image, we define a quantity that serves as a proxy for the signal-to-noise ratio (S/N) of every image and is intended to characterize the source visibility in a quantitative way. This quantity, which we denote as $\mathit{S/N}_\sigma$, is calculated in the following way. For a given image $\mathbf{x}$, the median $\tilde{\mathbf{x}}$ and standard deviation $\sigma_\mathbf{x}$ of the image are calculated using the sigma\_clipped\_stats method contained in the astropy package \cite{astropy}. Next, based on a sigma threshold, the image is segmented into two sections $\mathbf{x}_\mathrm{src}$ and $\mathbf{x}_\mathrm{bg}$, which are considered source and background regions, with the source region defined by

\begin{equation}
    \mathbf{x}_\mathrm{src} > \tilde{\mathbf{x}} + \tau \cdot \sigma_\mathbf{x}, \label{eq:sigma-mask}
\end{equation}
where $\tau$ is a threshold that is initially set to $\tau = 5$. For images where no source region is found, meaning no pixels of $\mathbf{x}$ satisfy Eq. \eqref{eq:sigma-mask}, the threshold $\tau$ is iteratively reduced by $0.5$ until a region can be identified. The background region is then defined by the remaining pixels in $\mathbf{x}$. Finally, the value for $\mathit{S/N}_\sigma$ is computed as the ratio of average pixel values in both regions, reading

\begin{equation}
    \mathit{S/N}_\sigma = \frac{\bar{\mathbf{x}}_\mathrm{src}}{\bar{\mathbf{x}}_\mathrm{bg}}. \label{eq:sigma-S/N}
\end{equation}
Under visual assessment, this quantity empirically shows to provide a good representation of how clearly a source is distinguished from background signals on the respective image. A few examples of this are shown in Fig. \ref{fig:sigma-S/N_examples}. We choose a threshold value of $\mathit{S/N}_\sigma = 5$ for our dataset, excluding all images that fall below this limit. Subsequently, we filter out images where the source is not well centered, or where the source is cropped by the cutout. This is done by excluding all images whose edge pixels have values exceeding a certain threshold, which we set to \num{0.8} times the maximum value of that image. Finally, we remove images with incomplete coverage, which are produced for sources that lie close to the edge of the sky area covered by the survey. In the end, this results in a dataset containing \num{106787} images of radio sources. A summary of the described selection cuts, with the corresponding numbers of excluded images at every step, is given in Table \ref{tab:selection_cuts}. Some examples of excluded images, and a selection of the images contained in the final selection that composes the LOFAR dataset, are shown in Fig. \ref{fig:excluded_examples} and Fig. \ref{fig:selection_examples}, respectively.\\

\begin{table}
    \caption{Numbers of removed and included images at each step of data selection.}
    \label{tab:selection_cuts}      
    \centering
    \begin{tabular}{l l l}
        \hline\hline
        Step     &   Removed   &   Total Selected   \\
        \hline
        cut-outs   &     0      & \num{314942} \\
        Broken    &  \num{1228}  & \num{313714} \\
        $\mathit{S/N}_\sigma$     & \num{201322} & \num{112392} \\
        Edge Pixels &  \num{5319}  & \num{107073} \\
        Incomplete  &  \num{286}   & \num{106787} \\
        \hline
        Total    & \num{208155} & \num{106787} \\
        \hline
    \end{tabular}
\end{table}

\begin{figure}
    \centering
    \resizebox{\hsize}{!}{\includegraphics{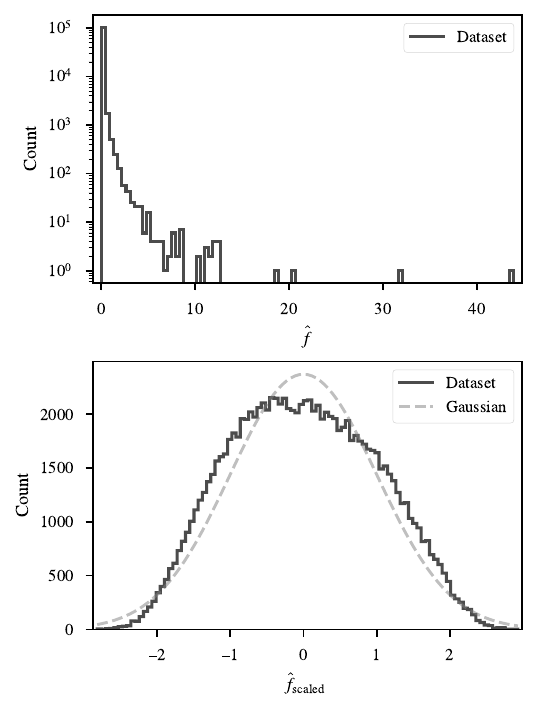}}
    \caption{Distributions of maximum flux values $\hat{f}$ (top) before and $\hat{f}_\mathrm{scaled}$ (bottom) after applying the Box-Cox transformation. The distributions comprise the final selection of the LOFAR dataset. For visualization, the bottom figure shows a standard normal distribution with an amplitude scaled to match the data distribution.}
    \label{fig:boxcox_distributions}
\end{figure}

Following the data selection, two steps of pre-processing are applied to the retrieved images. First, any negative pixel values of the images are set to \num{0}, since they are considered artifacts that result from the image reconstruction process and are physically unmotivated. Afterwards, every image is individually scaled to a range of $[0, 1]$ by applying the minmax-scaling according to

\begin{equation}
    x_\mathrm{scaled} = \frac{x - \min{\mathbf{x}}}{\max{\mathbf{x}} - \min{\mathbf{x}}}, \label{eq:minmax}
\end{equation}
where $x$ is any individual pixel of an image $\mathbf{x}$.
The scaled images are saved together with the original maximum pixel values for later retrieval and restoration of the original image. The radio images in the dataset show large variations in their maximum flux values, as can be contemplated in the top panel in Fig. \ref{fig:boxcox_distributions}. This property is desired to be reproduced when sampling from the trained model. However, this information is lost in the pre-processing, since the model is trained only on images that are individually scaled and therefore share the same magnitudes. Since the source flux can influence the image qualitatively even after scaling, e.g. in the value of $\mathit{S/N}_\sigma$, we choose to pass the maximum flux value of every image as an additional conditioning parameter during training. To standardize the distribution of those values, we apply a scaling known as Box-Cox power transform \cite{Box-Cox}, defined as 
\begin{equation}
    \hat{f}_\mathrm{scaled} = \frac{\hat{f}^\lambda - 1}{\lambda}, \label{eq:boxcox}
\end{equation}
where $\hat{f}$ and $\hat{f}_\mathrm{scaled}$ are the original and scaled maximum flux values, and $\lambda$ is a parameter that is optimized via maximum likelihood estimation to make the transformed distribution resemble a normal distribution, i.e. zero mean and  unit variance. This optimum is found for the LOFAR dataset at $\lambda = -0.23$. The transformed distribution is shown in the bottom panel of
Fig. \ref{fig:boxcox_distributions}.

\subsection{FIRST dataset}

\begin{figure}
    \centering
    \begin{subfigure}[t]{0.49\hsize}
        \centering
        \resizebox{\hsize}{!}{\includegraphics{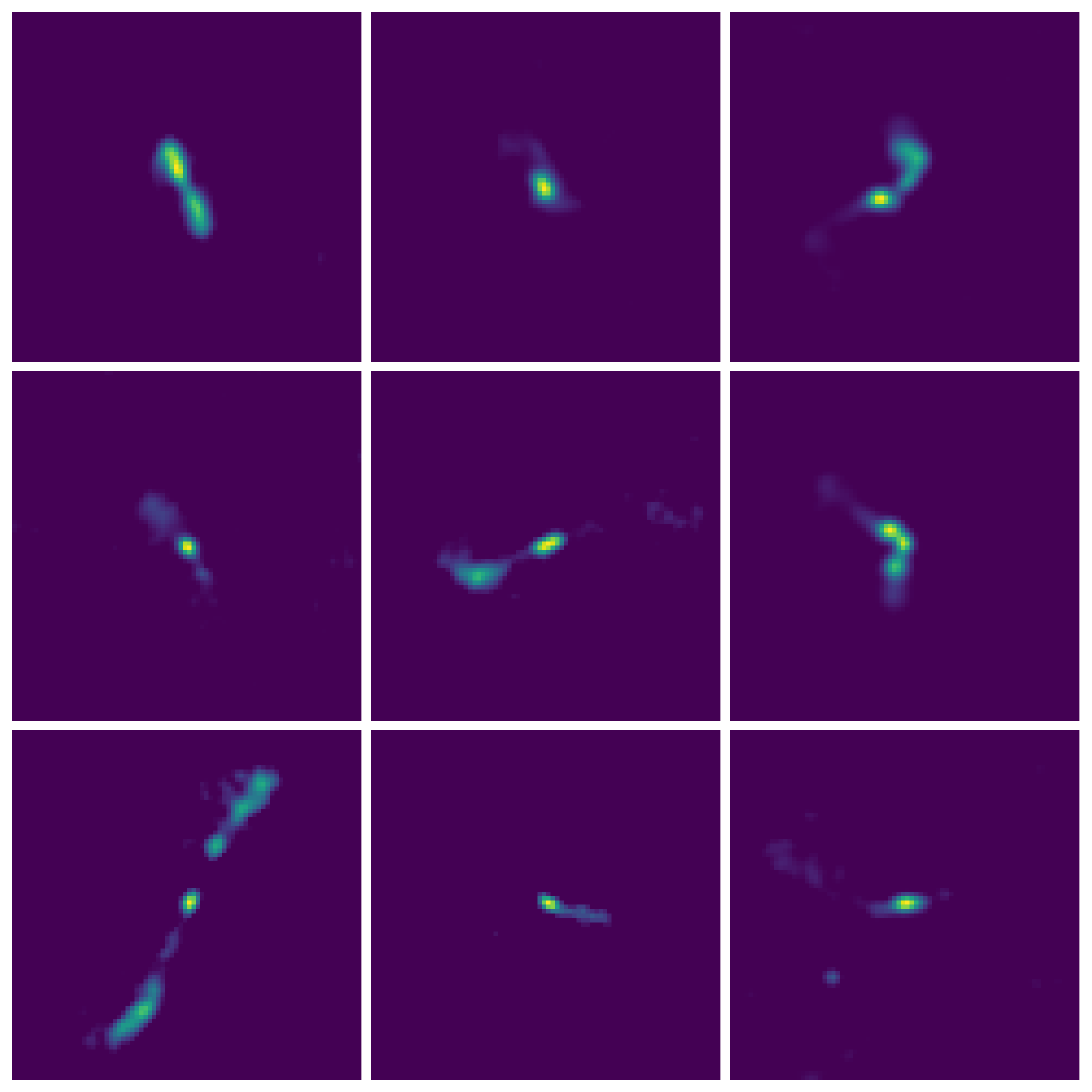}}
        \caption{FRI}
        \label{fig:sub:FIRST_FRI}
    \end{subfigure}
    \begin{subfigure}[t]{0.49\hsize}
        \centering
        \resizebox{\hsize}{!}{\includegraphics{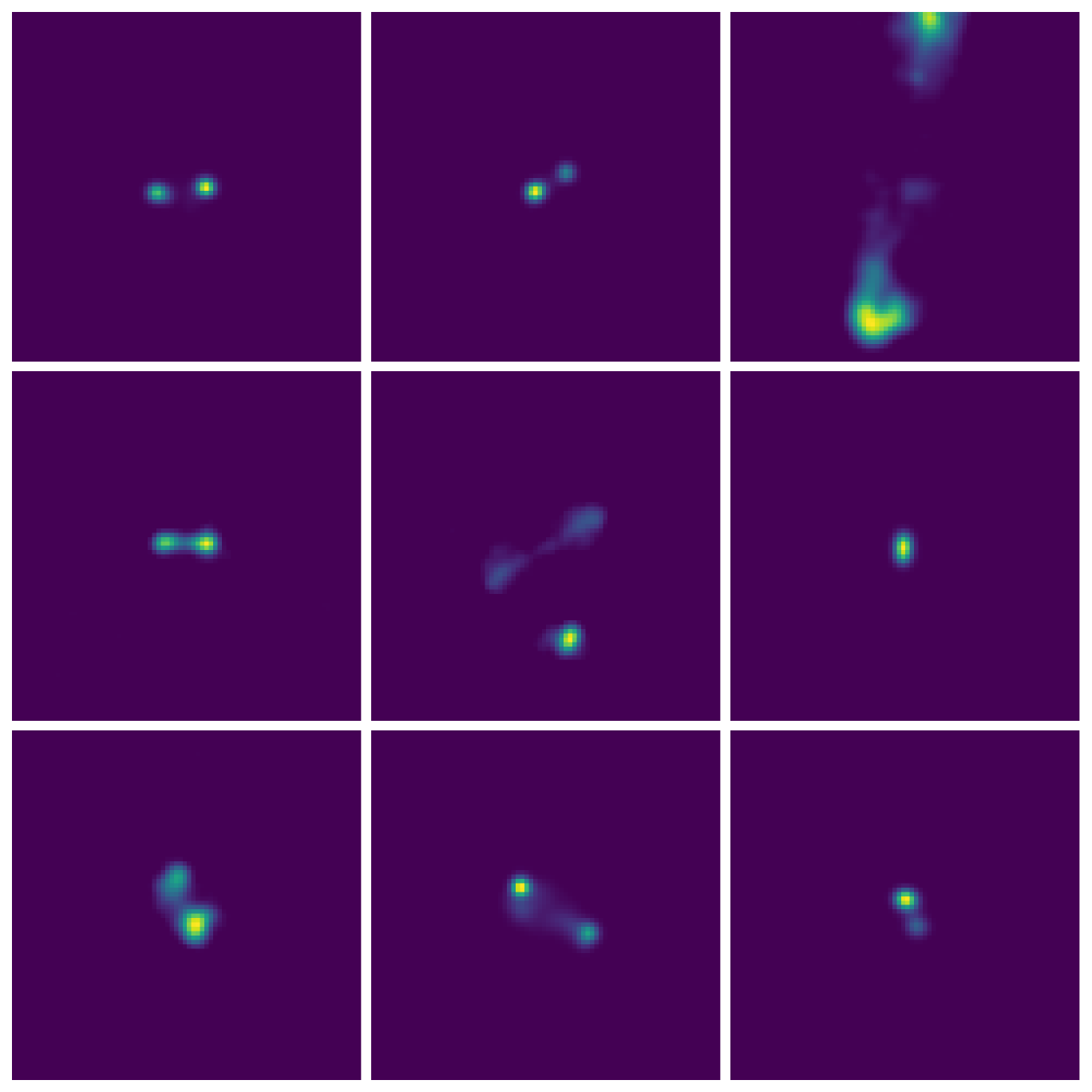}}
        \caption{FRII}
        \label{fig:sub:FIRST_FRII}
    \end{subfigure}
    \begin{subfigure}[t]{0.49\hsize}
        \centering
        \resizebox{\hsize}{!}{\includegraphics{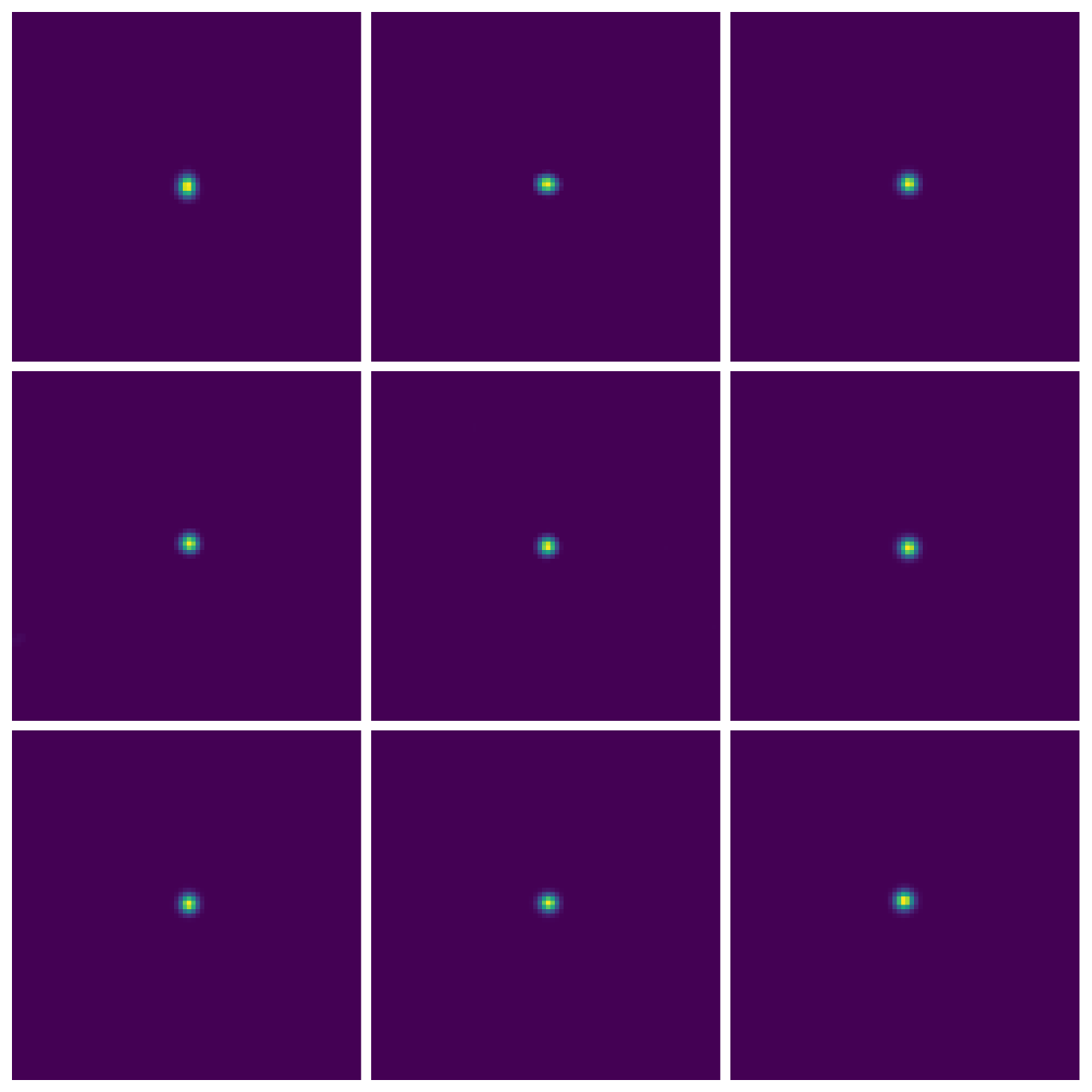}}
        \caption{Compact}
        \label{fig:sub:FIRST_Compact}
    \end{subfigure}
    \begin{subfigure}[t]{0.49\hsize}
        \centering
        \resizebox{\hsize}{!}{\includegraphics{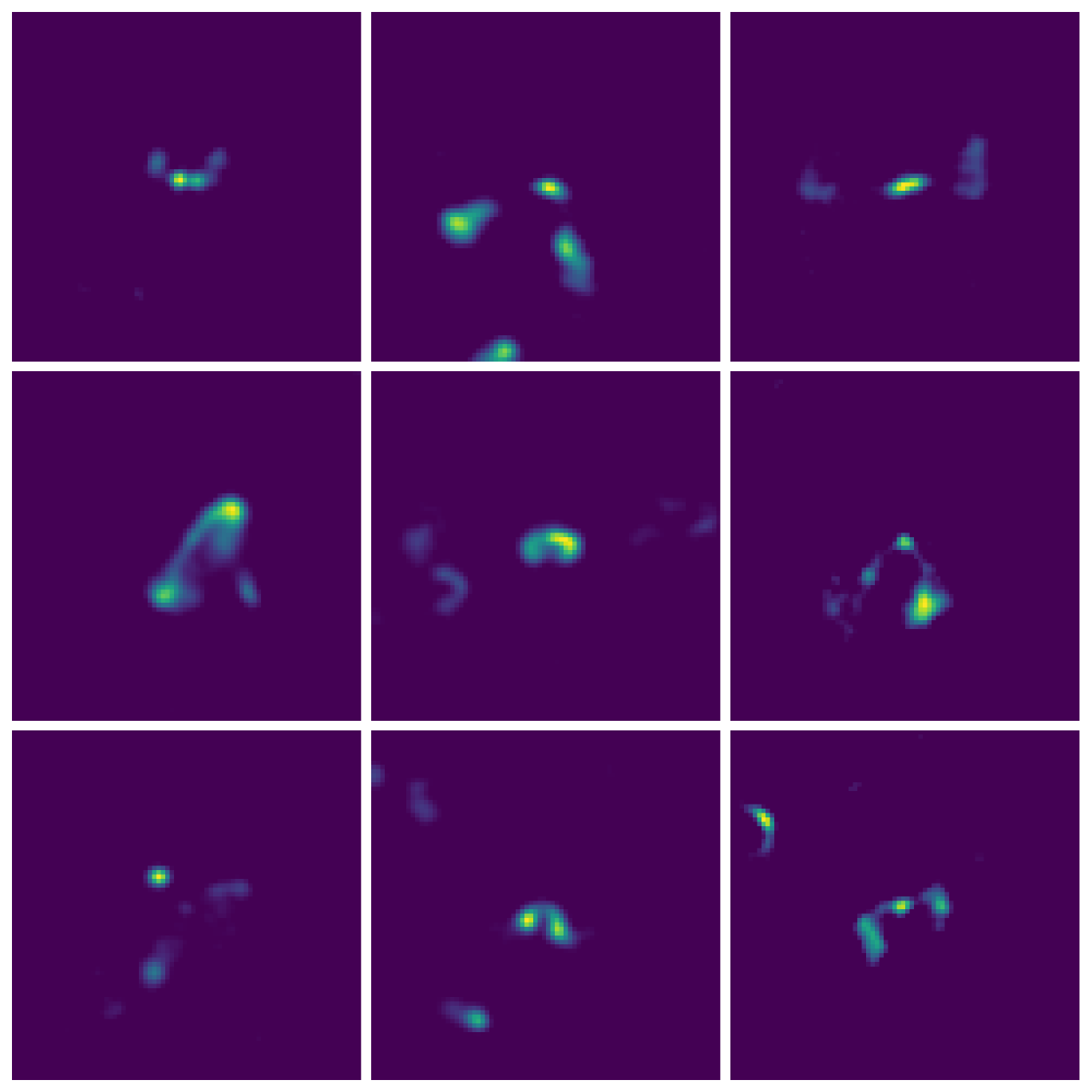}}
        \caption{Bent}
        \label{fig:sub:FIRST_Bent}
    \end{subfigure}
    \caption{Random examples of images from the FIRST dataset, showing one grid for each of the four morphological classes contained in the dataset, as indicated in the corresponding caption.}
    \label{fig:FIRST_Examples}
\end{figure}

A second, smaller dataset with morphological class labels is used to train a class-conditioned model, from hereon referred to as the FIRST model. This dataset was introduced in \cite{Griese+23} and contains \num{2158} images retrieved from the VLA survey of Faint Images of the Radio Sky at Twenty-Centimeters (FIRST), see \cite{FIRST_Survey}. Morphological class labels are collected from different catalogs and summarized into a classification scheme of four different categories. The first two correspond to the traditional classes from the Fanaroff-Riley (FR) scheme introduced in \cite{Fanaroff&Riley_1974}, with FR-I for extended sources that have their maximum radio emission close to the center, and FR-II for sources with maxima close to their edges. Further, unresolved point sources are classified as Compact, and finally sources for which the angle between the jets differs significantly from \ang{180} are classified as Bent. The numbers of examples for the different classes are given in Table \ref{tab:FIRST_classes}, a few examples of the images are shown in Fig. \ref{fig:FIRST_Examples}. As part of the pre-processing, the authors remove background noise by fixing all pixel values below three times the local RMS noise to that value, and subsequently applying minmax-scaling. For a more detailed description, we refer to \cite{Griese+23}.\\
The publicly available dataset \footnote{https://zenodo.org/records/7351724} provides the images in PNG format scaled to $[0, 255]$. Since the images are retrieved already individually scaled, we hence apply no conditioning on the flux values when training the FIRST model.
For our training, we crop the images from their original $300\!\times\!300$ pixels to $80\!\times\!80$ pixels around the image center, and subsequently rescale them to $[0, 1]$ according to Eq. \eqref{eq:minmax}. To train the FIRST model, for simplicity, we employ random undersampling to balance the dataset, limiting the number of images in every class to the size of the least populated one. This results in a reduced training set of \num{1392} images with equal amounts corresponding to the four classes.

\begin{table}
    \caption{Numbers of images corresponding to the different classes in the FIRST dataset.}
    \label{tab:FIRST_classes}
    \centering
    \begin{tabular}{l l l}
        \hline\hline
        Class     &   Images  &  Proportion\\
        \hline
        FRI  &  \num{495}  & \num{22.93}\%\\
        FRII  &  \num{925}  & \num{42.81}\%\\
        Compact  &  \num{391}  & \num{18.11}\%\\
        Bent  &  \num{348}  & \num{16.12}\%\\
        \hline
        Total & \num{2158}  & \num{100}\%\
    \end{tabular}
\end{table}

\section{Model and training} \label{sec:model-and-training}
In the following section, we describe technical details of the DM. This includes the neural network architecture, the training procedure and the sampling algorithm.

\subsection{Model architecture}

\begin{figure}
    \centering
    \resizebox{\hsize}{!}{\includegraphics{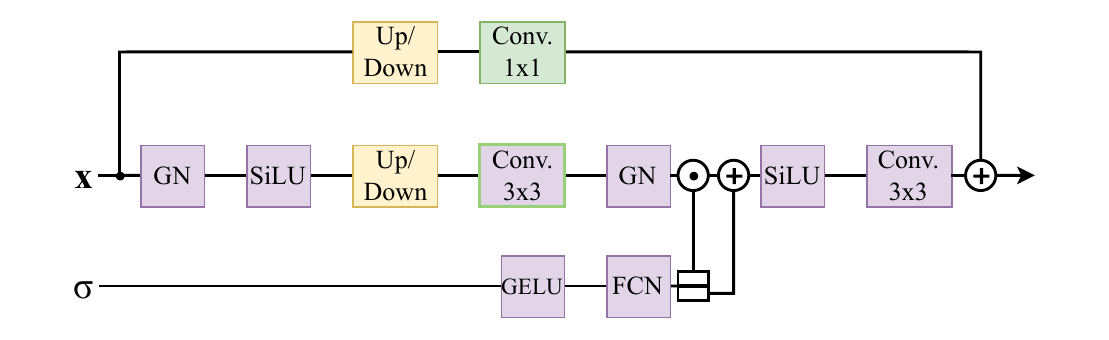}}
    \caption{Architecture of the U-Net building block. Optional elements of different variations are colored in agreement with Fig. \ref{fig:unet}. The first $3\!\times\!3$-convolution is always present. However, the included change in channel dimensions is optional.}
    \label{fig:unet_block}
\end{figure}

\begin{figure}
    \centering
    \resizebox{\hsize}{!}{\includegraphics{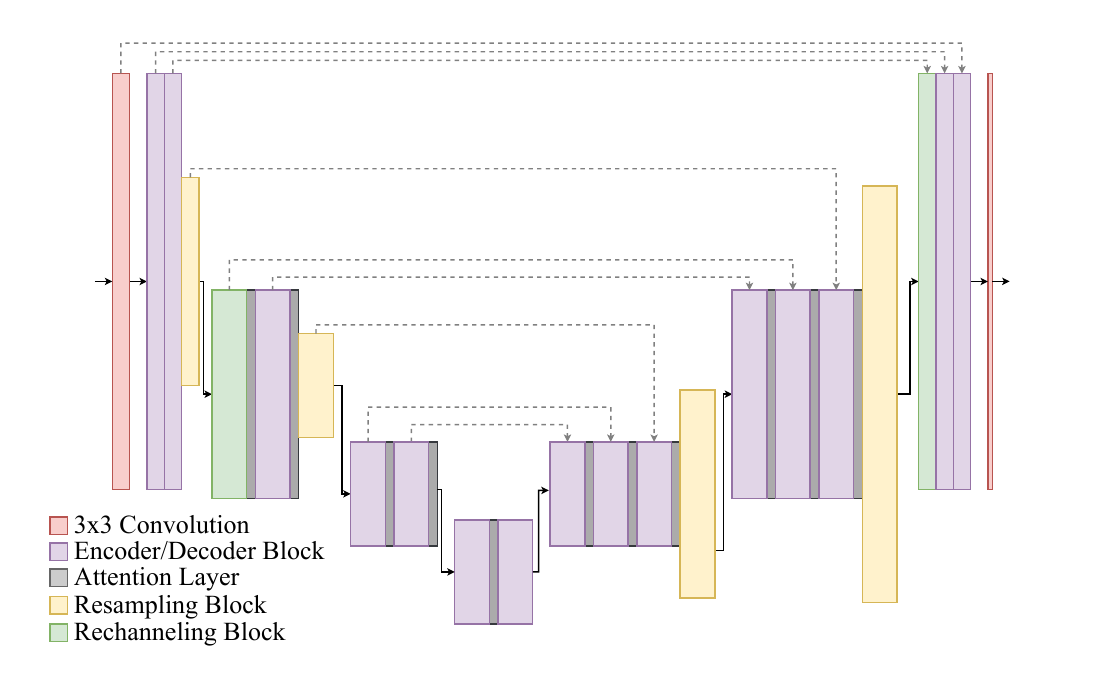}}
    \caption{Architecture of the U-Net. Skip connections between encoder and decoder are indicated with dashed lines.}
    \label{fig:unet}
\end{figure}

The denoiser neural network $D_\theta$ takes as input an image with added noise, additionally the corresponding noise level is injected as conditioning information throughout different parts of the network, together with the other optional conditioning information containing peak flux levels, or class labels respectively, for the two trained models. We refer to the entire conditioning information as context.\\

To implement the network, we employ a U-Net \citep{U-Net} architecture, which consists of an encoder and a decoder part with a bottleneck part in-between. The entire network architecture is illustrated in Fig. \ref{fig:unet}. Its constituents are mainly composed of convolutional neural network (CNN) layers, as well as self-attention layers. During a forward pass, the encoder gradually reduces the resolution of the feature maps while increasing the channel dimensions, the opposite happens in the decoder. In addition, the encoder and decoder are linked through skip connections at equal levels of resolution. All parts of the network are assembled using the same building block, which is illustrated in Fig. \ref{fig:unet_block}. Depending on the given function in different parts of the network, i.e. changing the resolution or number of channels, this block has slight variations in its architecture.\\

The standard block contains two successive sequences of group normalization (GN) \citep{GroupNorm}, Sigmoid Linear Unit (SiLU) activation \citep{SiLU_GELU} and a CNN layer with a kernel size of $3\!\times\!3$. The SiLU activation function is defined by
\begin{equation}
    \mathrm{SiLU}(x) = \frac{x}{1+\exp(-x)}.
\end{equation}
All elements are bypassed by an additional skip connection that is added to the output at the end of the block. For every single block in the entire network, the conditioning is applied after the second GN layer. The noise level is embedded using sinusoidal position encoding \citep{Vasvani+17} followed by a single-layer fully connected network (FCN). For the two different models, class labels are encoded with one-hot vectors and the peak flux values are passed as single neuron activations, respectively. This information is fed through an embedding network that consists of a two-layer FCN with Gaussian error linear unit (GELU) activation. The GELU activation is defined by
\begin{equation}
    \mathrm{GELU}(x) = x \cdot \mathrm{\Phi}(x),
\end{equation}
where $\mathrm{\Phi}(x)$ is the standard Gaussian cumulative distribution function. The advantage over the often used ReLU activation function is the smoother transition.
The embedding is subsequently added to the noise level embedding. The resulting vector is passed through another embedding network of identical architecture, and finally injected into the block through feature-wise linear modulation \citep{FiLM}. For some blocks, the first convolution layer changes the number of channels. In that case, a $1\!\times\!1$-kernel convolution layer is added to the skip connection to match the dimensions at recombination. For blocks that change the image resolution, down- or upsampling layers are added before the first convolution, as well as to the skip connection. Upsampling is done with the nearest neighbor algorithm, where each pixel is quadrupled. Downsampling is done by splitting the image into four equal parts, then applying $1\!\times\!1$-convolution with the four sub-images stacked along the input channel dimension.\\
In addition to the U-Net block, some parts of the network contain multi-head self-attention layers. We implement the efficient self-attention mechanism introduced in \citet{EffAttn}, preceded by a single GN layer. For an attention head with $c$ channels and an input feature map with $n$ pixels, three tensors $\mathbf{Q}$, $\mathbf{K}$ and $\mathbf{V}$ of dimensions $n\!\times\!c$ are obtained through a $3\!\times\!3$-CNN layer applied on the input. The attention tensor $\mathbf{A}$ is then computed as 

\begin{equation}
    \mathbf{A} = \mathrm{softmax}_1\left(\frac{\mathbf{Q}}{\sqrt{c}}\right)\cdot \mathrm{softmax}_2(\mathbf{K})^\mathrm{T}\mathbf{V}, \label{eq:attention}
\end{equation}
with the $\mathrm{softmax}_d$ function on a tensor, $\mathbf{T}$, defined as

\begin{equation}
    \mathrm{softmax}_d(\mathbf{T})^{(i_d)} = \frac{\exp(\mathbf{T})^{(i_d)}}{\sum_{j_d}\exp(\mathbf{T})^{(j_d)}}, \label{eq:softmax}
\end{equation}
where $d$ indicates one specific dimension of the tensor, $\mathbf{T}^{(i_d)}$ is the \textit{i}-th sub-tensor along that dimension, and the exponential function operates on the individual tensor components. This entire self-attention mechanism is carried out with multiple separate instances in parallel, which are called attention heads. The resulting attention tensors are subsequently concatenated and brought back to the input shape through another $3\!\times\!3$-CNN layer, followed by layer normalization \citep{layerNorm}, and finally added to the input via a skip connection.\\ 
Our implementation of the U-Net uses three levels of resolution, with each level containing two, respectively three resolution-preserving U-Net blocks for the encoder and decoder. Resampling blocks are added at the end of each resolution level, with exception of the lower-most in the encoder and upper-most in the decoder. The bottleneck part consists of two standard blocks with an attention layer in-between. On the lower two resolution levels, self-attention layers are added after all resolution-preserving blocks. The U-net is completed by skip connections at equal levels of resolution between blocks of the encoder and decoder, giving it the characteristic shape it is named after. The network is encompassed by two $3\!\times\!3$-convolutions at the beginning and end. Further technical details about different architectural choices are given in Table \ref{tab:U-Net}.\\

\begin{table}
    \caption{Details of the implemented U-Net architecture.}
    \label{tab:U-Net}      
    \centering
    \begin{tabular}{l l}
        \hline\hline  
        Parameter & Value \\    
        \hline                        
        Resolution levels & $(80, 40, 20)$ \\
        Initial channels & \num{128} \\
        Channel multipliers & $(1, 2, 2) $ \\
        Norm. groups & \num{32} \\
        Attention & (No, Yes, Yes) \\
        Attention heads & \num{2} \\
        Attention head channels & \num{32} \\
        Parameters & \\
        \hline
    \end{tabular}
    \tablefoot{
         Resolution levels indicate the image sizes in pixels on the different levels in the U-Net. The number of channels at different levels is given as a product of the initial channels and corresponding channel multiplier.
    }
\end{table}

To model the denoiser function, we employ a scaling to the network as described in \citet{Karras+22}, where the authors derive the following expressions from first principles to reduce variations in signal and gradient magnitudes. Let $F_\theta$ denote the neural network, then the parameterized denoiser function $D_\theta(\mathbf{x}; \sigma)$ as introduced in Sect. \ref{sec:diffusion-models} is defined as 
\begin{equation}
    D_\theta(\mathbf{x}; \sigma) = c_{\mathrm{skip }}(\sigma) \cdot \mathbf{x} + c_{\mathrm{out }}(\sigma) \cdot F_\theta\left(c_{\mathrm{in }}(\sigma) \mathbf{x} ; c_{\mathrm{noise}}(\sigma)\right), \label{eq:preconditioning}
\end{equation}
where the different constants are given by
\begin{align}
    &c_{\mathrm{skip }}(\sigma) = \sigma_\mathrm{data}^2 \cdot \left(\sigma^2+\sigma_\mathrm{data}^2\right)^{-1}, \label{eq:c_skip} \\
    &c_{\mathrm{out }}(\sigma) = \sigma \cdot \sigma_\mathrm{data} \cdot \left( \sigma^2+\sigma_\mathrm{data}^2\right)^{-\frac{1}{2}}, \label{eq:c_out} \\
    &c_{\mathrm{in }}(\sigma) = \left(\sigma^2+\sigma_\mathrm{data}^2\right)^{-\frac{1}{2}}, \label{eq:c_in} \\
    &c_{\mathrm{noise}}(\sigma) = \frac{1}{4} \ln{\sigma} \label{eq:c_noise},
\end{align}
with $\sigma_\mathrm{data} = 0.5$ fixed.

\subsection{Training}
An overview of all hyperparameters used for the different training runs is given in Table \ref{tab:hyperparameters}. As introduced in \citet{Karras+22}, noise levels $\sigma$ are sampled from a log-normal distribution referred to as $p_\mathrm{train}$, such that $\ln{\sigma} \sim \mathcal{N}\left(P_\mathrm{Mean}, P_\mathrm{Std}\right)$, where we find the parameter values given in Table \ref{tab:hyperparameters} to provide the best results. Following the same paper, we weigh the $L_2$ loss with $c_{\mathrm{out }}(\sigma)^{-2}$ (see Eq. \eqref{eq:c_out}) to keep a balanced loss magnitude across different noise levels. This results in a loss function $\mathcal{L}$ defined by
\begin{equation}
    \mathcal{L}(\theta) = \mathbb{E}_{\mathbf{x} \sim p_\mathrm{data}} \mathbb{E}_{\sigma \sim p_\mathrm{train}} \mathbb{E}_{\mathbf{n} \sim \mathcal{N}\left(\mathbf{0}, \sigma^2 \mathcal{I}\right)} \left[c_{\mathrm{out }}(\sigma)^{-2} \left\Vert D_\theta(\mathbf{x}+\mathbf{n} ; \sigma)-\mathbf{x}\right\Vert_2^2\right],
\end{equation}
which is used to optimize the model weights.\\
We train both models with the Adam optimizer \citep{adam}. For the LOFAR model, weights are stabilized by keeping an exponential moving average (EMA) of their values. During training, images are rescaled to $[-1, 1]$ using Eq. \eqref{eq:minmax}, with a subsequent rescaling step to obtain the desired interval boundaries. The model is trained with mixed precision, meaning forward pass and backpropagation are calculated with half-precision weights, but weight updates are done in full precision. In this context, we employ loss scaling to prevent underflow for small gradient values. For both the LOFAR and the FIRST model, the respective datasets are split into training and validation sets of 90\% and 10\% size, respectively. The latter is used to monitor the validation loss at regular intervals to ensure that the model is not overfitting the training data. To prevent this from happening, we applied dropout layers after each U-Net block. With the described architecture, the model was run with a total of about \num{3.1e8} trainable parameters. We employed a classical data augmentation by applying random horizontal and vertical flips, followed by a random rotation of multiples of $\ang{90}$. The training was performed using distributed data parallel training on two NVIDIA A100 GPUs and completed in approximately $\SI{19}{\hour}$ for the LOFAR dataset and $\SI{4}{\hour}$ for the FIRST dataset.\\

\begin{table}
    \caption{Hyperparameters used during training.}
    \label{tab:hyperparameters}      
    \centering
    \begin{tabular}{l l}
        \hline\hline  
        Parameter & Value\\    
        \hline                        
        Learning Rate & $\num{2e-5}$\\
        Batch Size & 256 (128) \\
        Iterations & \num{1e5} (\num{4e4}) \\
        EMA Rate & \num{0.9999} (-)\\
        $P_\mathrm{Mean}$ & \num{-2.5} \\
        $P_\mathrm{Std}$ & \num{1.8}\\
        Dropout Rate & \num{0.1} \\
        Context Dropout Rate & \num{0.1} \\
        \hline
    \end{tabular}
    \tablefoot{
         If present, numbers in brackets indicate the values used for training the FIRST model. Otherwise, the same values were used for both the training runs. EMA is only used for the LOFAR model.
    }
\end{table}

\subsection{Sampling} \label{sec:sub:Sampling}
In order to map a sample of pure Gaussian noise onto a realistic image, as described in Sect. \ref{sec:diffusion-models}, we solve Eq. \eqref{eq:scoreMatchedODE} numerically through discretization, using the neural network to evaluate the denoiser function. Our choices for the ODE solver, noise schedule $\sigma(t)$, and discretization time steps are taken from \cite{Karras+22}. We use the Heun's second-order method \citep{Ascher1998ComputerMF} as ODE solver, with the noise schedule defined as 
\begin{equation}
    \sigma(t) = t, \label{eq:noise_schedule}
\end{equation}
by which the noise level $\sigma$ and time parameter $t$ become interchangeable. Thus, Eq. \eqref{eq:scoreMatchedODE} simplifies to: 
\begin{equation}
    \frac{\mathrm{d} \mathbf{x}}{\mathrm{d} \sigma}=\frac{\mathbf{x} - D(\mathbf{x}; \sigma) }{\sigma}. \label{eq:ODE_final}
\end{equation}
The solution to Eq. \eqref{eq:ODE_final} can be understood as a trajectory through image space whose tangent vector points in the direction of the denoiser output for the respective image and noise level. In order to solve the equation numerically, we iteratively evaluate the right-hand side expression at previously selected discrete time steps, i.e. noise levels. Starting from pure Gaussian noise, we proceed to take small steps in the direction of the resulting vector to approximate the solution trajectory. The number of time steps $N$ presents a trade-off between sampling time and quality. For our purposes, we find $N =\num{25}$ time steps to be optimal. The noise levels $\{\sigma_\mathrm{i}\}$ for $ \mathrm{i} \in \{0, ..., N\}$ selected for the ODE solver are defined by

\begin{equation}
    \sigma_{1 \leq \mathrm{i} \leq N - 2}=
        \left(\sigma_{\max }{ }^{\frac{1}{\rho}}+\frac{i}{N-1}\left(\sigma_{\min }{ }^{\frac{1}{\rho}}-\sigma_{\max}{}^{\frac{1}{\rho}}\right)\right)^\rho,
\end{equation}
with $\sigma_0 = \sigma_{\max}$, $\sigma_{N-1} = \sigma_{\min}$ and $\sigma_N = 0$. The values given to the free parameters are $\sigma_{\max}=80$, $\sigma_{\min}=\num{2e-3}$ and $\rho = 7$. The detailed procedure used for sampling is described in algorithm \ref{alg:Sampling}. Conditioned sampling is realized by using the linear combination $\Tilde{D}_\theta$ as described in Sect. \ref{sec:sub:guided_diffusion}, where the guidance strength $\omega$ varies for different parts of this work. After sampling, pixel values are clamped to $[-1, 1]$, and rescaled to $[0, 1]$ for further analysis.\\
For a fixed model, the wall-clock inference time is generally bound by the capacities of the underlying hardware. Since the denoiser network is called at every time step during the sampling process, its total duration grows  linearly with the number of time steps, $N$. The inference time further depends on the model size and the choice of the solver. For instance, note that due to the 2\textsuperscript{nd} order correction in algorithm \ref{alg:Sampling}, the model is run twice for every sampling step. On a single Nvidia A100 GPU, with our choice of $N=25$, it takes $\SI{3}{\minute}\, \SI{17}{\second}$ to sample a batch of 1000 images.\\

\begin{algorithm}
    \caption{Sampling algorithm, given denoiser neural network $D_\theta(\mathbf{x};\sigma)$ and noise levels $\sigma_{i}$ with $i \in \{0, ..., N\}$}
    \label{alg:Sampling}
    \begin{algorithmic}[1]
        \doublespacing
        \Procedure{HeunSampler}{$D_\theta(\mathbf{x};\sigma), \; \{\sigma_i\}_{i \in \{0, ..., N\}}$} 
            \LeftComment{Seed image from pure Gaussian noise:}
            \State \textbf{sample} $\mathbf{x}_0 \sim \mathcal{N}\left(\mathbf{0}, \sigma_0^2 \mathbb{I} \right)$
            \For{$i = 0, ..., N-1$} \Comment Discretization steps
                \State $\displaystyle \mathbf{d}_i \gets \frac{\mathbf{x}_i - D_\theta(\mathbf{x}_i; \sigma_i) }{\sigma_i}$  \Comment Evaluate ODE at $\sigma_i$
                \State $\mathbf{x}_{i + 1} \gets \mathbf{x}_i + (\sigma_{i+1} - \sigma_i) \mathbf{d}_i$ \Comment Euler step
                \LeftComment{Apply 2\textsuperscript{nd} order correction if not $\sigma_i=0$:}
                \If{$i+1 \neq N$}
                    \LeftComment{Evaluate ODE at $\sigma_{i+1}$:}
                    \State $\displaystyle \mathbf{d}'_i \gets \frac{\mathbf{x}_{i+1} - D_\theta(\mathbf{x}_{i+1}; \sigma_{i+1}) }{\sigma_{i+1}}$ 
                    \LeftComment{Trapezoidal rule:}
                    \State $\mathbf{x}_{i + 1} \gets \mathbf{x}_i + (\sigma_{i+1} - \sigma_i) \frac{1}{2} (\mathbf{d}_i + \mathbf{d}'_i)$ 
                \EndIf{}
            \EndFor{}
            \State \Return{$\mathbf{x}_N$} \Comment Final image
        \EndProcedure{}
    \end{algorithmic}
\end{algorithm}

\section{Evaluation metrics} 
\label{sec:evaluation-metrics}

When assessing the quality of generated images resulting from our training, two separate aspects are to be considered: In general, the images should resemble the training dataset in terms of signal intensity and source morphology. In addition, the different types of conditioning information used during model training should be accurately represented in the samples, indicating that the corresponding image properties can be controlled through those parameters. In this section we describe different metrics used to monitor both of these aspects.

\subsection{Image quality} 
\label{sec:sub:image_quality_metrics}

\begin{figure}
    \centering
    \resizebox{\hsize}{!}{\includegraphics{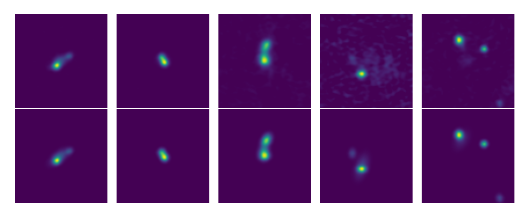}}
    \caption{Examples of images from the LOFAR dataset (top), with their corresponding reconstructed flux model images generated with PyBDSF (bottom).}
    \label{fig:pybdsf_examples}
\end{figure}

\begin{figure}
    \centering
    \resizebox{\hsize}{!}{\includegraphics{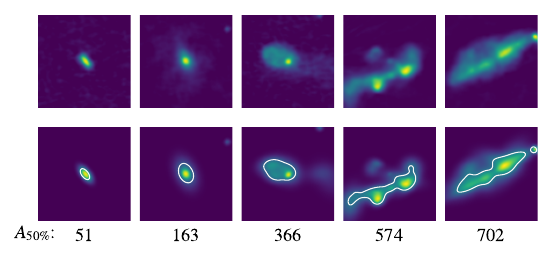}}
    \caption{Examples of images from the LOFAR dataset (top) and flux model (bottom), with increasing values of $A_{50\%}$ from left to right. The bottom figures indicate the calculated area with white contours, the corresponding $A_{50\%}$ values are shown below.}
    \label{fig:A50_examples}
\end{figure}

In order to compare a set of generated images to the training dataset, we calculate the distributions of different single-image quantities and compare the resulting histograms for the two datasets. This is done for the mean value and standard deviation of the scaled images. In addition, we compare the distributions of pixel values for both datasets, i.e. here one data point corresponds to a single pixel rather than an entire image.\\
To further compare quantities that are more descriptive of the source properties, we run the Python Blob Detector and Source Finder (PyBDSF) \citep{pyBDSF} on both datasets to obtain a corresponding flux model image for every sample. This allows us to characterize the sources in an automated way and analyze their properties separately from the background. The settings used for PyBDSF are fixed for all images, with the values  given in Table \ref{tab:pyBDSF_settings}. A few examples of training images and the corresponding source model are shown in Fig. \ref{fig:pybdsf_examples}. As a result of the PyBDSF analysis, we obtain a value of integrated source model flux for every image, which again is used to compare the distributions between datasets. Finally, as an indicator of the source extension, we calculate the minimum area $A_{50\%}$ that contains 50\% of the integrated model flux, given in pixels. Under visual inspection, this quantity correlates well with the perceived extension of the source, as demonstrated in Fig. \ref{fig:A50_examples}. We note that this is not equivalent to its actual physical size. For instance, a single source with two bright, small and widely separated hot-spots has a large physical extension; however, the $A_{50\%}$ area will be small, since only the constrained regions of high signal intensity will be covered. Therefore, this quantity should be regarded as a property of the image, rather than the source itself.

\begin{table}
    \caption{Settings used for the PyBDSF analysis.}
    \label{tab:pyBDSF_settings}      
    \centering
    \begin{tabular}{l l}
        \hline\hline  
        Parameter & Value\\    
        \hline                        
        thresh\_isl & 5\\
        thresh\_pix & \num{0.5} \\
        mean\_map & "Const" \\
        rms\_map & True \\
        thresh & "Hard" \\
        \hline
    \end{tabular}
\end{table}

\subsection{Flux conditioning} \label{sec:sub:flux_conditioning_metrics}
\begin{figure}
    \centering
    \resizebox{\hsize}{!}{\includegraphics{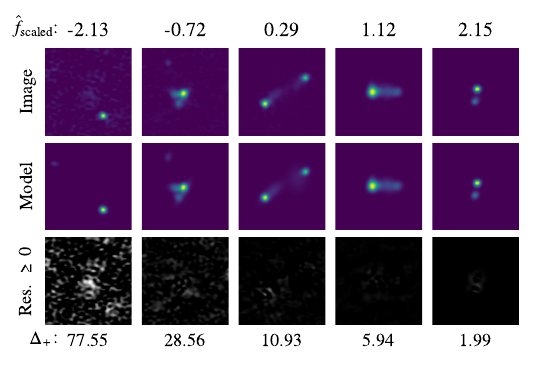}}
    \caption{Examples of positive residuals for different values of $\hat{f}_\mathrm{scaled}$, increasing from left to right. The top row shows the original images from the LOFAR dataset, the middle row shows the flux model, and the bottom row shows the positive residual flux. For the latter, the color values are scaled between 0 and \num{0.116}, which is the maximum value of the depicted residuals. Values of $\hat{f}_\mathrm{scaled}$ are indicated on top and values of $\Delta_+$ on the bottom.}
    \label{fig:pybdsf_residual_plot}
\end{figure}

For the LOFAR model, as described in Sect. \ref{sec:training-data}, we condition the model on the standardized peak flux $\hat{f}_\mathrm{scaled}$ to preserve flux information that is otherwise lost through the rescaling of pixel values. Typically, images with lower peak flux show more prominent background noise, since the source is not as bright in comparison. We quantify this property by calculating the residual with respect to the PyBDSF source model. Since this model ideally only describes the source and not the background, the difference should be representative of the background seen in the image. To further reduce the influence of residuals caused by the limited accuracy of flux modeling, we consider only the positive residual values. This positive flux residual $\Delta_+$ is then given by
\begin{equation}
    \Delta_+ = \sum_i (x_i - \Tilde{x}_i) \cdot \Theta(x_i - \Tilde{x}_i), \label{eq:residual}
\end{equation}
where $i$ runs over all image pixels $x_i$ and model image pixels $\Tilde{x}_i$, and $\Theta$ is the Heaviside step function. A few examples are shown in Fig. \ref{fig:pybdsf_residual_plot}, showcasing the relationship between $\hat{f}_\mathrm{scaled}$ and $\Delta_+$. To evaluate whether the conditioning information is learned correctly, we quantitatively compare the relation between the two properties for the training images and generated samples.

\subsection{Class conditioning}
Finally, in order to evaluate how well the morphological information contained in the class labels is learned by the FIRST model, we train a classifier on the FIRST dataset and compare its performance between the training data and the class-conditioned samples.\\
The classifier is composed of a CNN-based encoder part and a subsequent FCN-head that maps the encoder output onto one of the four classes. The encoder is implemented using the resnet-18 architecture \citep{He+15}, a residual CNN that takes an image as input and outputs a vector of 512 dimensions. The head consists of two fully connected layers with 256 hidden units and a four-dimensional output vector representing the classification logits for the four classes. We use batch norm after the encoder and apply dropout after the encoder and after the linear layers of the head, for which we make no use of non-linearities.\\

We train the model for 200 epochs with the cross-entropy loss on the predicted class labels, using the Adam optimizer \citep{adam} with a learning rate scheduler (LRS) that reduces the learning rate by a given factor when learning stagnates after a given number of epochs, called patience. All training hyperparameters are given in Table \ref{tab:classifier_hyperparameters}. During training, we apply a set of random augmentations consisting of random blurring, addition of random Gaussian noise, random cropping, random rotations and random vertical flips. The blurring is done with a $3\!\times\!3$ Gaussian kernel, where the standard deviation is sampled uniformly between \num{0.1} and 1. Similarly, the added noise has a standard deviation uniformly sampeled between \num{1e-4} and \num{0.1}. The size of the crop is chosen uniformly between \num{0.9} and 1 times the original image size, and the crop is subsequently resampled to the original image size. The rotations are performed with an arbitrary random angle, where resulting gaps in the image are filled with pixels of value \num{-1}. For the classifier training, the entire dataset of \num{2158} images is separated into a training (\num{1758} images), test, and validation (200 images each) set, which approximately corresponds to proportions of \num{0.9}, \num{0.1} and \num{0.1}. We employ loss weighting as balancing strategy, meaning for every image the training loss is weighted with the reciprocal proportion of the corresponding class in the dataset.\\
With the trained classifier, we compare its performance on the test split and on the generated data set. For this, we evaluate the overall classification accuracy and calculate the confusion matrices to obtain a detailed overview of the accuracy for the individual classes.

\begin{table}
    \caption{Hyperparameters used for the classifier training.}
    \label{tab:classifier_hyperparameters}      
    \centering
    \begin{tabular}{l l}
        \hline\hline  
        Parameter & Value\\    
        \hline                        
        Learning Rate & $\num{1e-3}$\\
        Batch Size & 128 \\
        Epochs & 200 \\
        LRS Factor & \num{0.135}\\
        LRS Patience & 50 \\
        LRS Min. Learning Rate & \num{1e-5}\\
        Dropout Rate & \num{0.5} \\
        \hline
    \end{tabular}
\end{table}

\section{Results} 
\label{sec:results}

\subsection{Image quality} \label{sec:sub:image_quality_results}
\begin{figure}
    \centering
    \resizebox{\hsize}{!}{\includegraphics{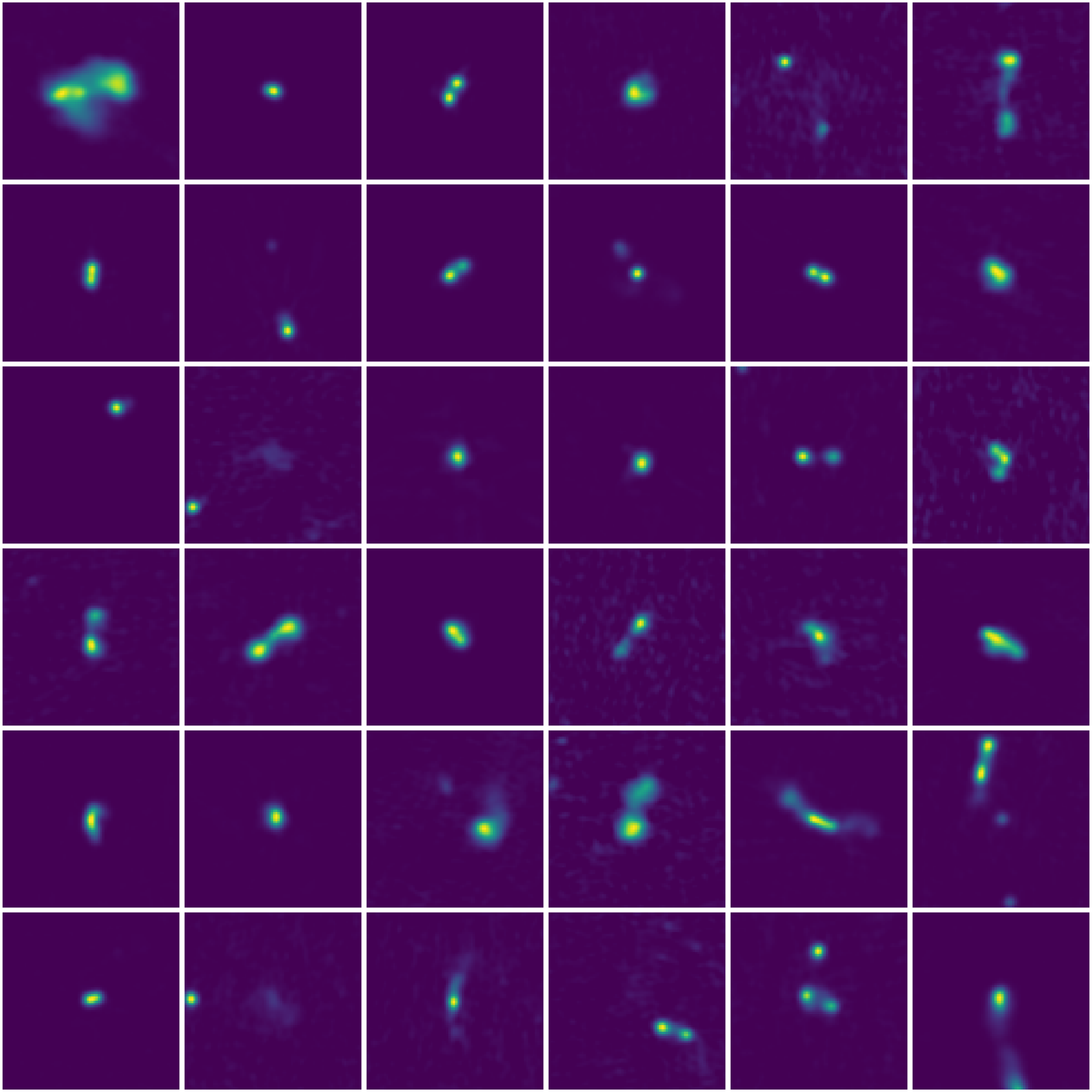}}
    \caption{Random examples of images generated with the trained LOFAR model.}
    \label{fig:res:LOFAR_samples}
\end{figure}

To evaluate the LOFAR model, we sample a set of \num{50000} images after training. This requires setting the conditioning parameter $\hat{f}_\mathrm{scaled}$ as a sampling input to every image. To resemble the properties of the training dataset in this regard, we model the distribution of $\hat{f}_\mathrm{scaled}$ shown in the bottom plot of Fig.~\ref{fig:boxcox_distributions} and sample the conditioning parameters from this model distribution. This is done by first choosing one of \num{100} histogram bins with probability proportional to the bin counts, and then uniformly sampling a value within the bin boundaries. To produce images conditioned on those values, we set the guidance strength to $\omega = 0.1$.\\
A selection of images generated with the LOFAR model is shown in Fig. \ref{fig:res:LOFAR_samples}. A first visual comparison between Fig. \ref{fig:selection_examples} and Fig.~\ref{fig:res:LOFAR_samples} shows that the samples are not distinguishable by eye from the training images. This is confirmed by further visual comparison of subsets larger than the image grids provided in this work. To illustrate the similarity between the generated samples and training data, we select a few generated samples and identify the five most similar images from the training dataset for each. This is shown in Appendix \ref{sec:app:nearest-neighbors}. We proceed to assess the sample quality in a more quantitative manner.\\

\begin{figure}
    \includegraphics[width=\hsize]{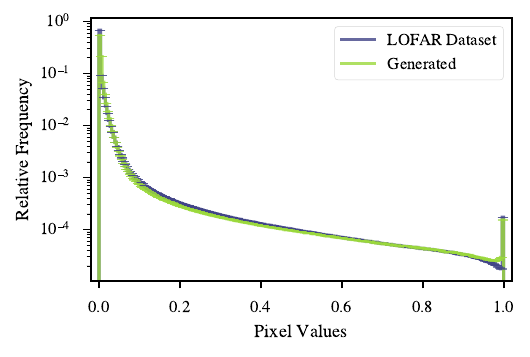}
    \caption{Pixel intensity distribution for the LOFAR dataset and LOFAR model-generated images.}
    \label{fig:res:pixel_distr}
\end{figure}

\begin{figure}
    \centering
    \includegraphics[width=\hsize]{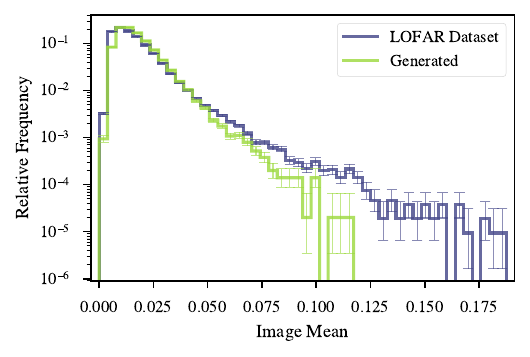}
    \caption{Image mean value distribution for the LOFAR dataset and LOFAR model-generated images.}
    \label{fig:res:mean_distr}
\end{figure}

\begin{figure}
    \centering
    \includegraphics[width=\linewidth]{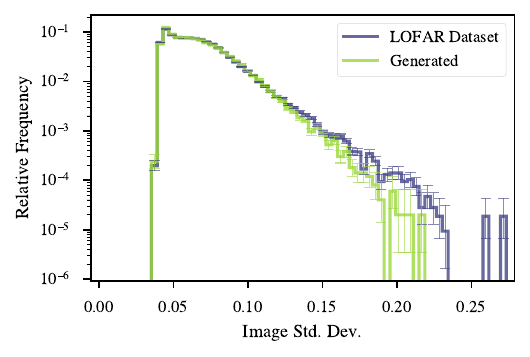}
    \caption{Image standard deviation distribution for the LOFAR dataset and LOFAR model-generated images.}
    \label{fig:res:sigma_distr}
\end{figure}

\begin{figure}
    \centering
    \includegraphics[width=\hsize]{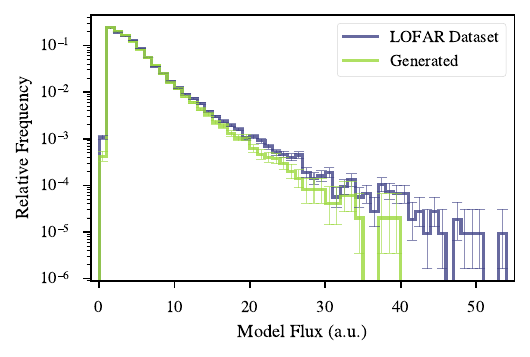}
    \caption{BDSF model flux distribution for the LOFAR dataset and LOFAR model generated-images.}
    \label{fig:res:bdsf_flux}
\end{figure}

\begin{figure}
    \centering
    \includegraphics[width=\hsize]{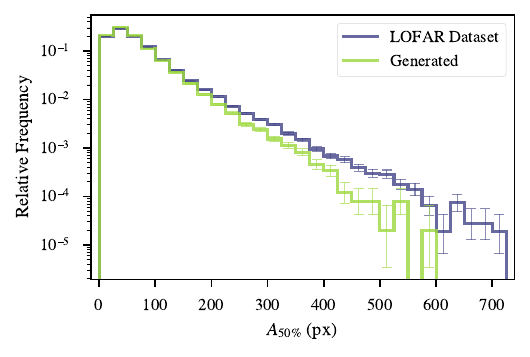}
    \caption{BDSF 50\%-Area $A_\mathrm{50\%}$ distribution for the LOFAR dataset and LOFAR model-generated images.}
    \label{fig:res:bdsf_A50}
\end{figure}

The distributions described in Sect. \ref{sec:sub:image_quality_metrics} are shown in Figs. \ref{fig:res:pixel_distr} to \ref{fig:res:bdsf_A50}, calculated over both the LOFAR dataset and the generated images. Error margins correspond to the $1\sigma$ frequentist confidence intervals \footnote{Calculated using astropy's stats.poisson\_conf\_interval \citep{astropy}.}.\\
The pixel value distribution shown in Fig.~\ref{fig:res:pixel_distr} is very well replicated by the sampled images. Only the sharp dip at the right end of the distribution, close to pixel values of \num{1}, appears more blunt in the sample distribution. We note that the outlier at the highest-value bin in the training distribution is an artifact of scaling the pixel values to $[0, 1]$, i.e. every single image has exactly one pixel of value 1. This is not explicitly implemented in the model training, hence the steepness of this feature is not expected to be replicated by the sampling process that is inherently noisy. The peak is only replicated by clamping the pixel values to a maximum of 1 after the sampling procedure. The same explanation holds for the lowest-value bin, for which pixels are slightly underrepresented in the generated images. Apart from the described details, the pixel value distribution is matched to a high degree.\\

The distributions of image mean and standard deviation in Fig.~\ref{fig:res:mean_distr} and Fig.~\ref{fig:res:sigma_distr} show high compatibility between real and generated data in lower pixel value ranges, whereas the counts in the upper ranges shift towards lower values for the generated images with increasing relative deviations. This indicates a subtle bias of the model towards images of lower pixel activity. The generated images also show fewer occurrences for very small mean values, which is likely a direct consequence of the underrepresented small pixel values seen in Fig.~\ref{fig:res:pixel_distr}.\\
The deviation towards higher values is also observed in the distribution of model flux in Fig.~\ref{fig:res:bdsf_flux} and $A_{50\%}$ in Fig.~\ref{fig:res:bdsf_A50}, both quantities obtained from the pyBDSF analysis. Altogether, those observations indicate that the model produces slightly fewer images with large source extensions as compared to the training data set. Apart from that, the shapes of the distributions are generally very well reproduced, indicating that the properties of the generated images are realistic and representative of the training dataset.\\

\subsection{Flux conditioning}
\begin{figure}
    \centering
    \includegraphics[width=\hsize]{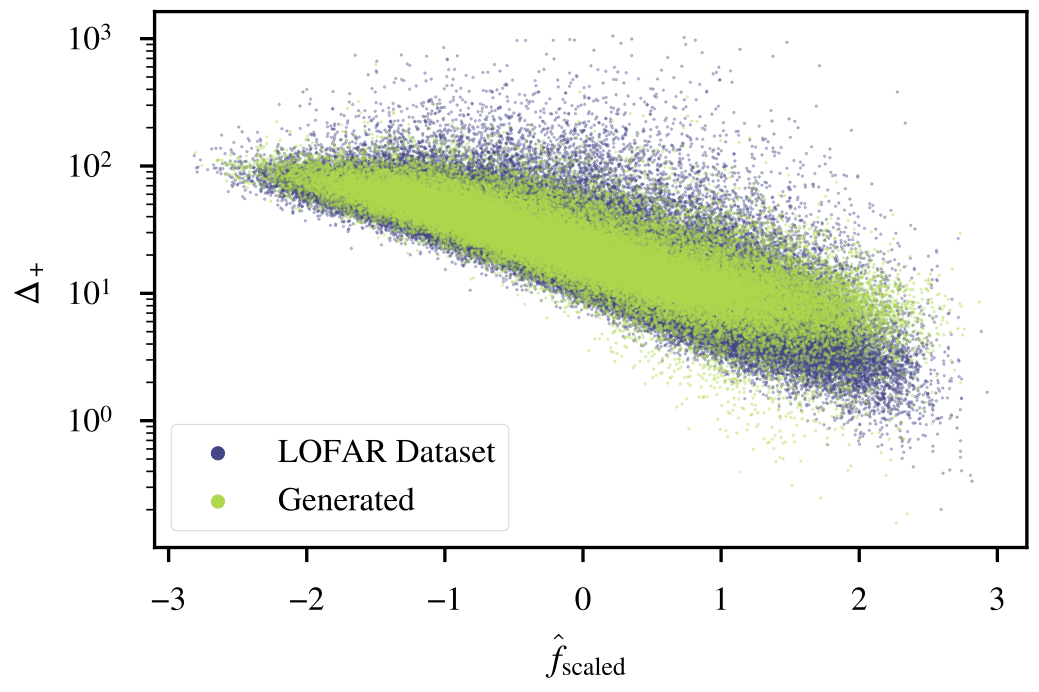}
    \caption{Scatter plot showing the negative correlation between the positive flux model residual $\Delta_+$ and the scaled peak flux, $\hat{f}_\mathrm{scaled}$, for the LOFAR dataset and LOFAR model generated images.}
    \label{fig:res:residual_vs_flux}
\end{figure}

Next, we investigate how well the information contained in the peak flux conditioning is learned by the model. As discussed in Sect. \ref{sec:sub:flux_conditioning_metrics}, we expect a negative correlation between the positive flux model residual $\Delta_+$ and the scaled peak flux, $\hat{f}_\mathrm{scaled}$. Figure~\ref{fig:res:residual_vs_flux} shows this relation as a scatter plot for the LOFAR data set and the set of generated images, where each data point corresponds to one image. The expected negative correlation is confirmed for the training data set, and proves to be well replicated by the generated images, meaning the encoded information is successfully learned during training. This happens especially well for the lower end of the distribution of  $\hat{f}_\mathrm{scaled}$, where both scatter plots are almost identical. At higher levels, there seems to be a cut-off at around $\Delta_+ \sim 5$, below which the population of generated images becomes sparser. With higher levels of $\hat{f}_\mathrm{scaled}$, the background signals in the training images become increasingly weak, and possibly the information becomes too nuanced to be efficiently learned during training. Also, this effect might be related to the difficulties of the model in replicating pixel values that are very close to 0. Furthermore, the training data shows more outliers towards higher values of $\Delta_+$ over the entire range. This is explained by instances of sources with extended, diffuse emission, which the pyBDSF model struggles to accurately represent as part of the source. The lack of those in the generated images can be attributed to the under-representation of extended sources established in Sect. \ref{sec:sub:image_quality_results}. Nonetheless, this result clearly indicates that the conditioning mechanism for the peak flux value works as intended.

\subsection{Class conditioning}
\begin{figure}
    \centering
    \begin{subfigure}[t]{0.49\hsize}
        \centering
        \resizebox{\hsize}{!}{\includegraphics{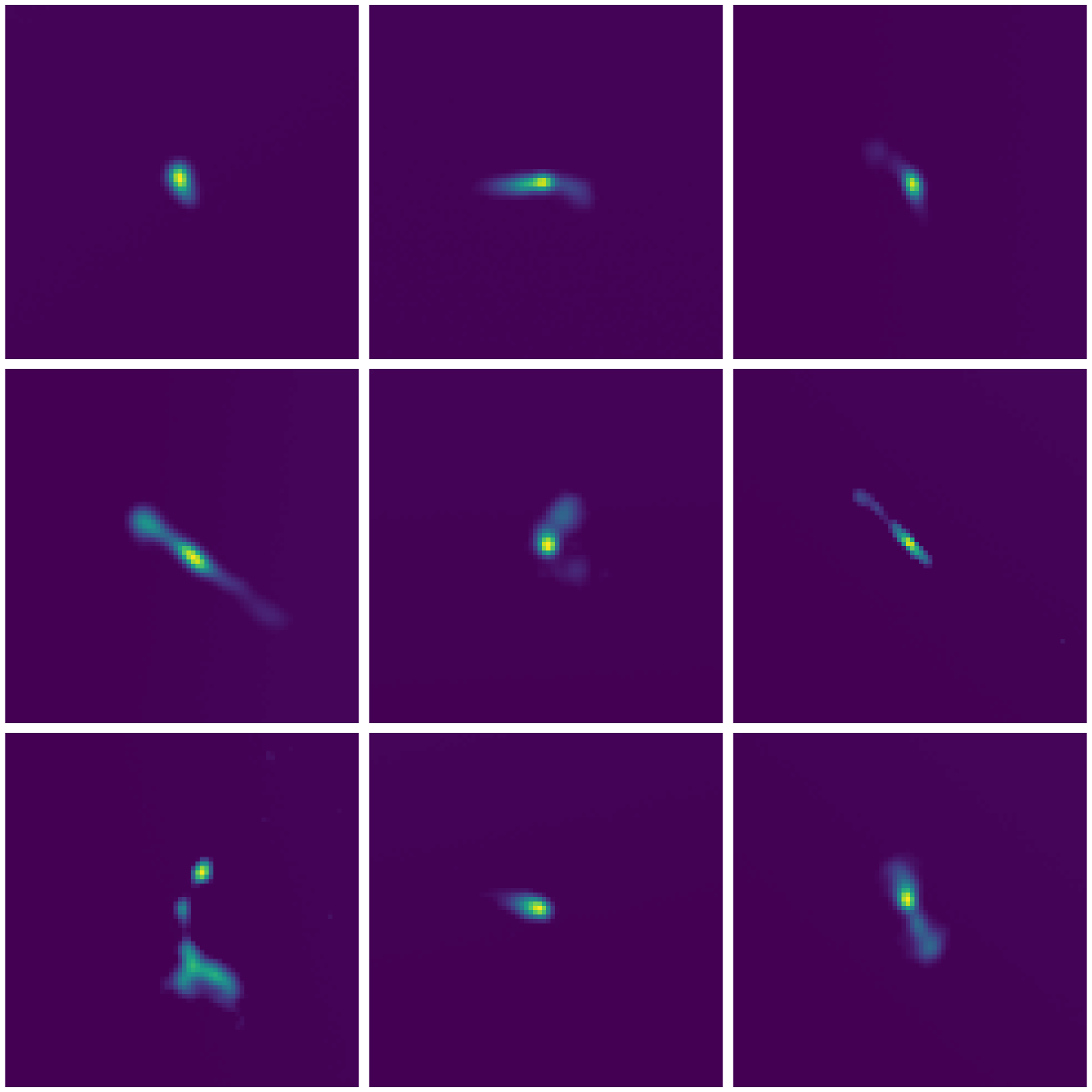}}
        \caption{FRI}
        \label{fig:sub:samples_FRI}
    \end{subfigure}
    \begin{subfigure}[t]{0.49\hsize}
        \centering
        \resizebox{\hsize}{!}{\includegraphics{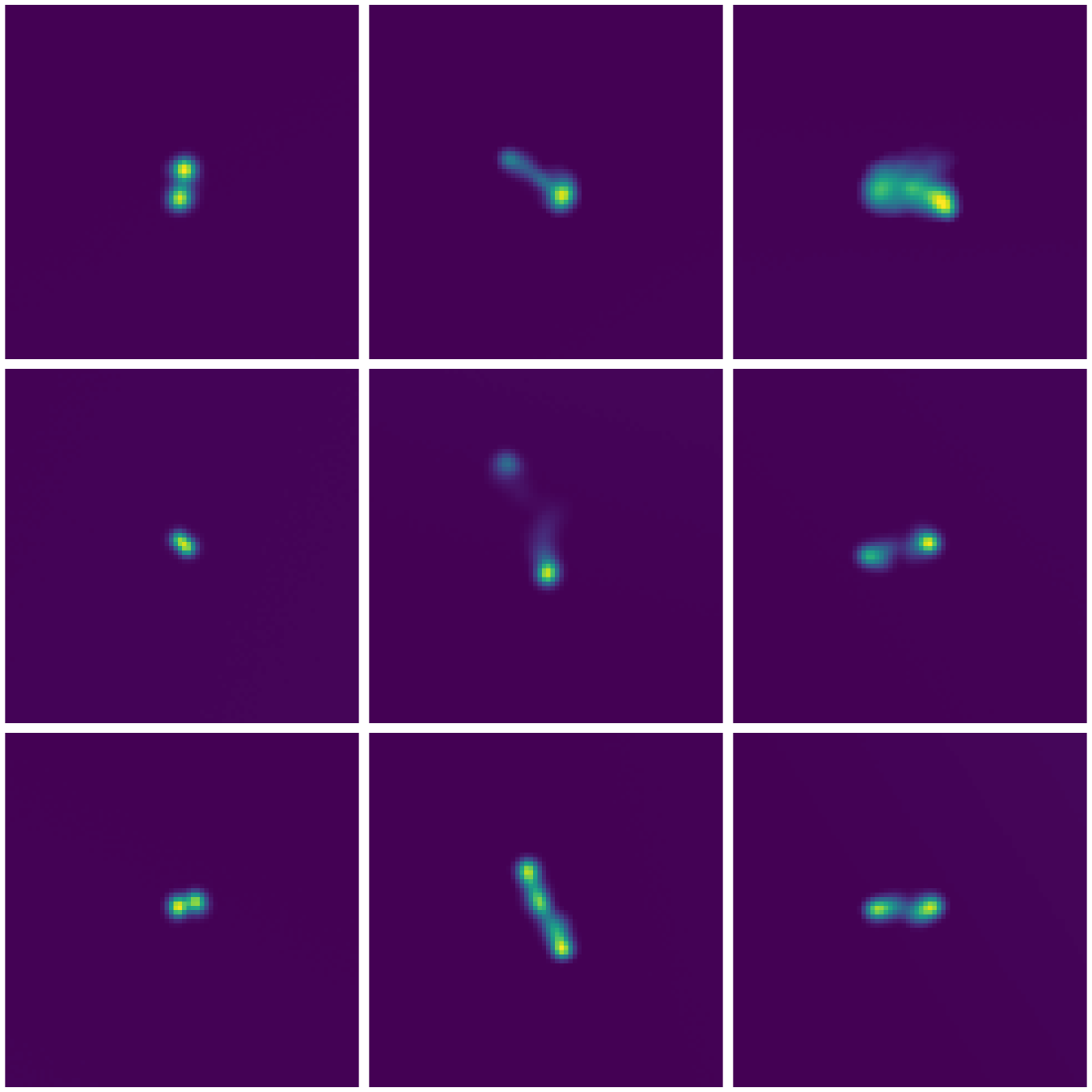}}
        \caption{FRII}
        \label{fig:sub:samples_FRII}
    \end{subfigure}
    \begin{subfigure}[t]{0.49\hsize}
        \centering
        \resizebox{\hsize}{!}{\includegraphics{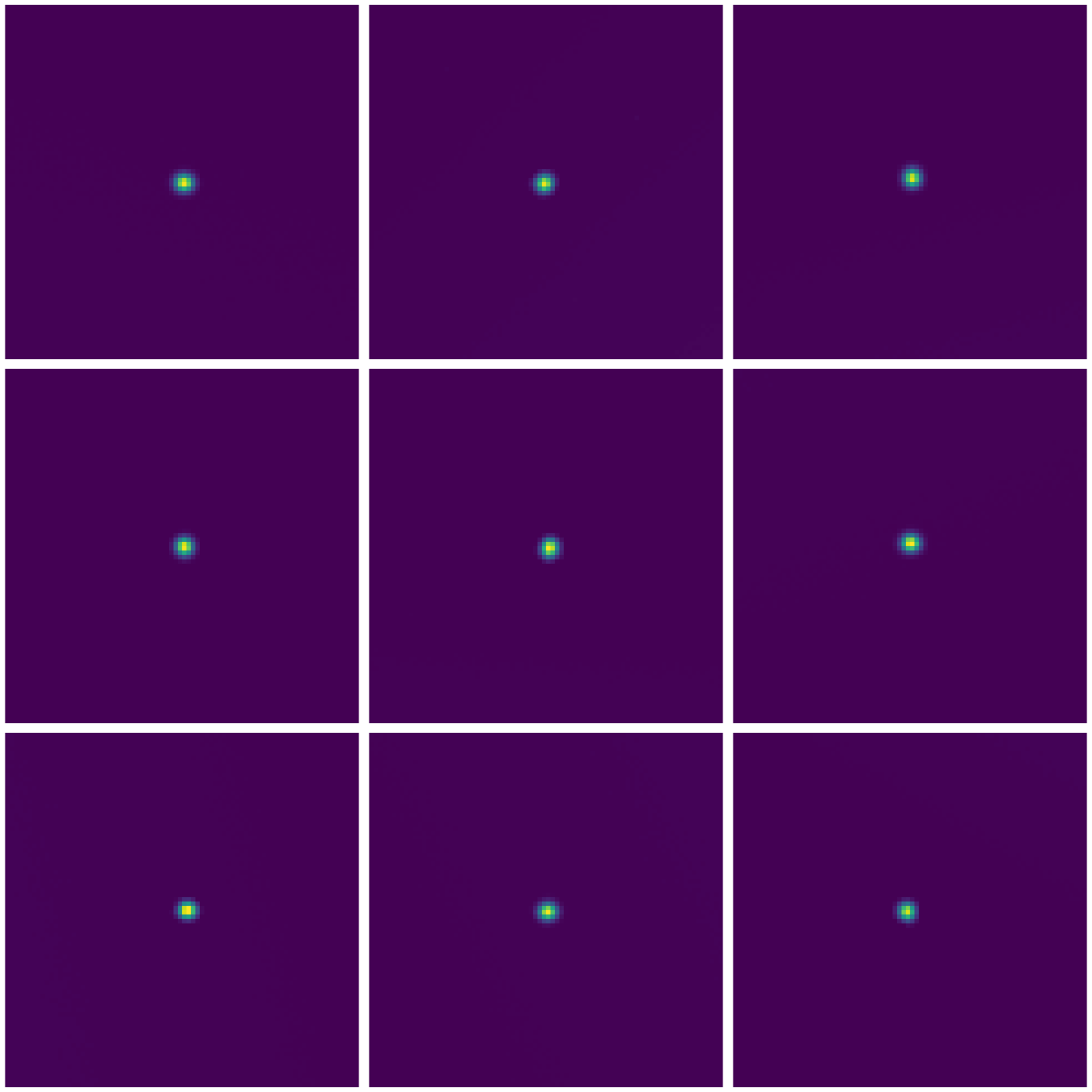}}
        \caption{Compact}
        \label{fig:sub:samples_Compact}
    \end{subfigure}
    \begin{subfigure}[t]{0.49\hsize}
        \centering
        \resizebox{\hsize}{!}{\includegraphics{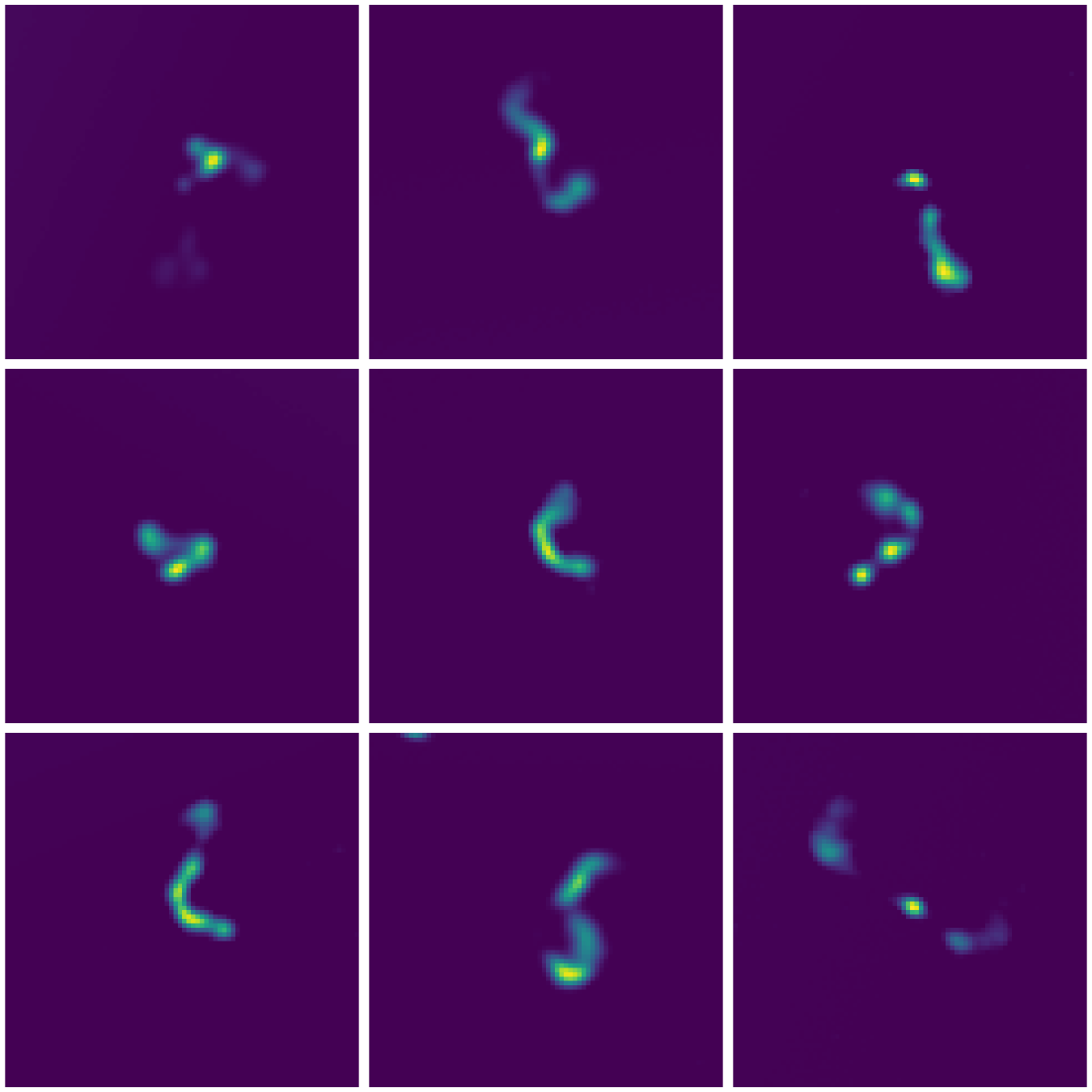}}
        \caption{Bent}
        \label{fig:sub:samples_Bent}
    \end{subfigure}
    \caption{Random examples of images generated with the FIRST model, showing one grid for each of the four morphological classes contained in the dataset, as indicated in the corresponding caption.}
    \label{fig:res:FIRST_Samples}
\end{figure}

We proceed to evaluate the class-conditioning mechanism of the FIRST model. For this, we sample a set of \num{2000} images for each of the four morphological classes using the trained model, experimenting with different values of guidance strength $\omega$. We obtain the best results with $\omega = 0.75$. Figure~\ref{fig:res:FIRST_Samples} shows a random selection of those images, which again under visual inspection provide an accurate representation of the training dataset.\\

\begin{figure}
    \centering
    \begin{subfigure}[t]{\hsize}
        \centering
        \resizebox{\hsize}{!}{\includegraphics{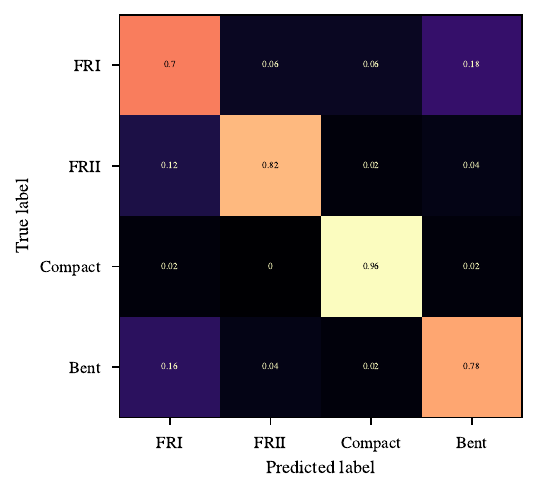}}
        \caption{FIRST Dataset}
        \label{fig:sub:Classifier_FIRST}
    \end{subfigure}
    \begin{subfigure}[t]{\hsize}
        \centering
        \resizebox{\hsize}{!}{\includegraphics{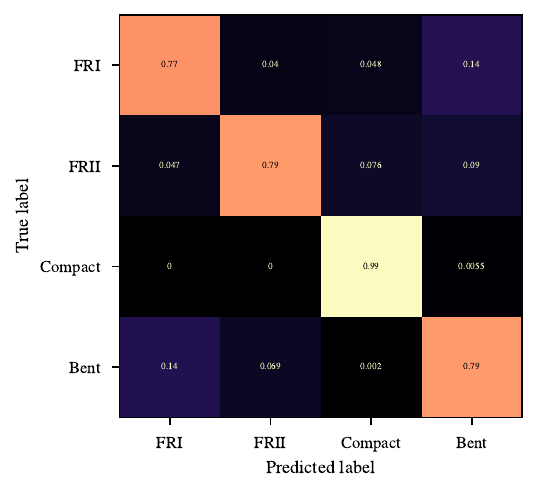}}
        \caption{Generated Images}
        \label{fig:sub:Classifier_Generated}
    \end{subfigure}
    \caption{Confusion matrices of the classifier trained on the FIRST dataset, for the test split (top) and the generated images (bottom).}
    \label{fig:res:Classifier_results}
\end{figure}

The classifier trained on the FIRST dataset achieves an accuracy of \num{81.5}\% over the test split, the corresponding confusion matrix is shown in the top panel of Fig.~\ref{fig:res:Classifier_results}. The classifier performs quite well over all classes. It shows most difficulties to distinguish between classes FRI and Bent in both ways, and also a significant proportion of FRIIs is mislabeled as FRIs. The rest of the sources is classified with high accuracy. Remarkably, the total accuracy is slightly improved over the set of generated images, reaching a value of \num{83.4}\%. While the bidirectional confusions between FRIs and Bent-tail sources are still present, the mislabeling of FRIs for FRIIs disappears. Further experiments reveal that the accuracy does decrease for samples generated with lower values of guidance strength, results are shown in Table \ref{tab:guidance_strength_accuracies}. This relation is expected, since at lower $\omega$ the sample fidelity is reduced in exchange for higher variability, effectively resulting in more overlap between the conditioned distributions in image space for the different classes. 

\begin{table}
    \caption{Classifier accuracy for sets of images sampled with different values of guidance strength, $\omega$.}
    \label{tab:guidance_strength_accuracies}      
    \centering
    \begin{tabular}{l l}
        \hline\hline  
        $\omega$ & Accuracy\\    
        \hline                        
        0.1 & 0.735\\
        0.2 & 0.747\\
        0.5 & 0.801\\
        0.75 & 0.834\\
        \hline
    \end{tabular}
\end{table}

\section{Discussion} 
\label{sec:discussion}

In this work, we have shown the use of DMs for synthetic generation of high-quality, realistic images of radio galaxies. Our model is capable of generating images whose properties resemble real images of radio galaxies. This is shown by the accurate replication of metric distributions that characterize different properties of the imaged sources and confirmed through visual comparison. In addition, the model is able to learn and accurately reproduce detailed morphological characteristics, which is demonstrated by the stable classifier accuracy over the class-conditioned samples. Furthermore, our results on, both, peak-flux and class conditioning show that a DM can be trained with different kinds of additional parameters that relate to any characteristic of the image, such as properties of the background or morphological features, and reliably recreate those relations at the sampling stage. This allows for precise control over the properties of simulated images, which can be tailored depending on the scientific use case for the synthetic data.\\

The ability to realistically reproduce fine-grained morphological properties of radio sources in a targeted, controlled way is critical for the generation of synthetic large-survey sky maps, where current methods of generation either employ simplified source models of composite 2D Gaussians that are limited in their amount of detail or use classically augmented real observations, which can lead to imbalances and lack of diversity. As we show in our work, the use of a DM can mitigate both problems, leading to significant improvements in the quality of simulated data. In addition, with our peak flux conditioning mechanism, this can be accurately done over a large dynamic range, which is relevant for simulating large-scale maps. \\

Our methodology represents a step forward with respect to previous work regarding DMs in radio astronomy. We replace the framework of "discrete-time" DMs and move on to training a "continuous-time" DM by following the methods proposed in \cite{Karras+22}. This allows us to reduce the number of denoising steps to 25 during sampling, which constitutes an improvement of at least one order of magnitude in comparison to previous work with DMs, which employ between \num{250} \citep{Drozdova+2024} and \num{1000} \citep{Sortino+2023} steps. Even though the sampling algorithm used here employs two model passes per sampling step, this still constitutes a significant reduction of the neural network evaluations required for image generation, which directly translates into a reduction of computing needs, while still achieving high quality results. Nonetheless, as is the case with DMs in general, the sampling time elevated through significant computational requirements constitutes the principal limitation to the use of our model. Further improvements in this regard might be made through a detailed exploration of different architectural choices. The possibility to reduce the number of model parameters while maintaining sample quality would facilitate a further decrease in both training and sampling time. Especially the latter would make the use of the model more accessible to the scientific community, reducing the extent of required computational resources. This could be approached in a simple way by experimenting with different parameters of the network employed in this work or, more profoundly, by implementing architectural improvements to the U-Net proposed in \cite{Karras+23} intended to optimize training and model efficiency. Another interesting approach is the idea of consistency models \citep{Song+23}, which are trained to solve the PF-ODE in a single inference step. These can be trained stand-alone or through knowledge distillation from an already trained DM, where our model can be employed. A different, complementary improvement would be the use of latent diffusion \citep{Rombach+21},  which could potentially reduce the computational cost, while  allowing for a sampling of larger images. \citet{Sortino+2023} demonstrated promising results in this regard.\\

In addition, the ability to sample from a DM with control over the morphological class is novel in the field of radio astronomical imaging. While the RADiff model introduced in \cite{Sortino+2023} offers control over the source shape through conditioning on segmentation masks, this does not influence the distribution of signal intensity over the emitting regions, which is the main property relevant for the Fanaroff-Riley classification scheme. At this point, we achieve high sample fidelity with respect to classifier accuracy using a guidance strength of $\omega = 7.5$, higher than the typical values of $\omega \approx 0.1 \text{ to } 0.2$ used for natural image sampling \citep{Ho&Salimans22}. This confirms the expected relation that a higher value of $\omega$ leads to a higher sample fidelity. Consequently, it raises the question to what extent the model is limited in the mode coverage of the different classes, and how the sample variety is affected by this. Some insight could be gained by implementing metrics typically used in the field of natural image generation; for instance, Fréchnet inception distance (FID) or inception score (IS), which compare real and generated datasets based on different quantities derived from network activations of a pre-trained image classifier that for this use should be specific to radio astronomical images. In the case of FID, such an approach is used in \cite{Sortino+2023}. Although a precise quantitative assessment of the meaning and reliability of such results is challenging, the potential understandings can prove useful for the further development of class-conditioned sampling. We leave this exercise for future endeavors. Overall, we also expect both the class-conditioned DM and the classifier used for evaluation to largely benefit from a larger dataset of labeled sources, which in our case is two orders of magnitude smaller than the unlabeled LOFAR dataset. A minor improvement can potentially be made here with a different balancing strategy for the DM training by employing loss weighting instead of undersampling, as done for the classifier training.\\

Another limitation of our method is derived from the fact that it is trained on radio images that have been observed, cleaned and deconvolved with a limited variety of telescope configurations and a fixed processing pipeline. Hence, the models capacity of simulating realistic thermal or systematic noise, associated with said properties in the training data, is accordingly constrained in its variability. This also applies to the clean beam of the telescope, which varies across observations. Inversely, some of the low surface brightness features in the DM simulated radio images are caused by clean artifacts which the DM emulates from the training data.

Furthermore, our work presents a scheme for selection of image data derived from the LoTSS-DR2 that is versatile for adaptation to individual purposes, making it universally applicable for extracting training datasets suitable for different machine learning applications. The introduced $\mathit{S/N}_\sigma$ presents a heuristic that allows for a selection of images where the source morphologies are visible with different degrees of clarity; thus, a variation in the selection cut threshold allows us to achieve a trade-off between completeness and purity. This can also be leveraged to tailor the DM to a specific need; namely, setting a lower threshold to entirely represent properties of the telescope and imaging process, versus a higher threshold to focus on source morphologies as in our work. The parameter can also be used as conditioning information for free post-training control of the generated images with respect to this property. This can in fact be done with any other parameter available in the source catalog, which is seamlessly integrated in the pre-processing pipeline. Alternative ways of quantifying source visibility might include the use of image segmentation models and improved estimates of noise and artifacts and could potentially lead to more precise selection cuts and result in larger datasets of useful images. Our data selection procedure can be expanded to extract cut-outs of larger size, thereby including more extended or badly centered sources that would otherwise not fit on the cut-out and be excluded due to activated edge pixels. Cut-outs larger than the actual images would also facilitate the use of random rotations at arbitrary angles as a method of classical data augmentation, removing the need to fill up gaps in the rotated image with artificial pixel values, which is undesirable for the training of generative models. The missing pixels would instead be completed with the true image data available in the larger cut-outs.\\

Our DM can be applied to different use cases in radio astronomy. \cite{Rustige+2023} investigated the use of GAN-generated images for augmenting a dataset used to train a classifier. In their work, although it does fulfill various benchmarks of sample quality and fidelity, the generated images still show minor differences from the training data noticeable by eye. While simple classifiers experience a benefit from the data augmentation, this is not observed for more complex architectures. With the increased quality of the samples produced from our work, it is worth exploring whether those results can be improved and to what extent different machine learning models can generally benefit from data augmentation with generated images.\\

Furthermore, sets of single generated source images can be assembled to produce synthetic large-scale survey maps. For this, an artificial radio sky can be populated by sampling from well-known distributions of sky sources, such as the two-point correlation function of radio galaxies observed in the National Radio Astronomy Observatory (NRAO) VLA Sky Survey, as described in \cite{Song&Schwarz16}. The maps can then be realized by extracting sources from generated images and adding them on a background map that can either be real, generated from an analytical model, or, in a more extensive approach, may be sampled from a generative model trained on source-free regions only. This approach would require us to train the DM on a set of images free of noise and major artifacts to exclusively simulate the source emission. Such a dataset can be obtained through threshold-based masking of high-S/N images, reliably isolating the source signal. Careful consideration should however be given to potential effects caused by selection biases. Noise induced through the measurement and analysis procedure could then be added to the synthetic map through a subsequent step, using such software as the LOFAR simulation tool\footnote{\url{https://github.com/darafferty/losito}} \citep{LoSiTo}, which simulates the measurement result for a given sky model and observation configuration.\\

The design and training of our DM can be extended to serve further purposes. While we are currently working with a single image channel, this is not a necessary constraint, since, by construction, our model supports an arbitrary number of channels. With the use of corresponding training data, this can be leveraged to produce consistent images in different bands typical for radio observations or even to experiment with simultaneous generation of images in entirely different energy domains. Another potential use case is doing super-resolution of radio images similar to the work of \cite{Reddy+24}, which might prove useful for classification tasks where different source morphologies become hard to distinguish at the limit of resolution. Overall, this work provides an important step forward in the ability to synthesize realistic radio galaxy image data in a precise and controlled way.

\section{Data availability}
The code implemented in this work is available at \url{https://github.com/tmartinezML/LOFAR-Diffusion}.\\

\begin{acknowledgements}
We thank the anonymous referee for a very constructive report.
MB and TVM acknowledge support by the Deutsche Forschungsgemeinschaft under Germany’s Excellence Strategy – EXC 2121 Quantum Universe – 390833306 and via the KISS consortium (05D23GU4) funded by the German Federal
Ministry of Education and Research BMBF in the ErUM-Data action plan.\\

\end{acknowledgements}

\bibliographystyle{aa}
\bibliography{references}

\begin{appendix}
\section{Samples with the highest similarity training images} \label{sec:app:nearest-neighbors}

\begin{figure*}
    \centering
    \includegraphics[width=17cm]{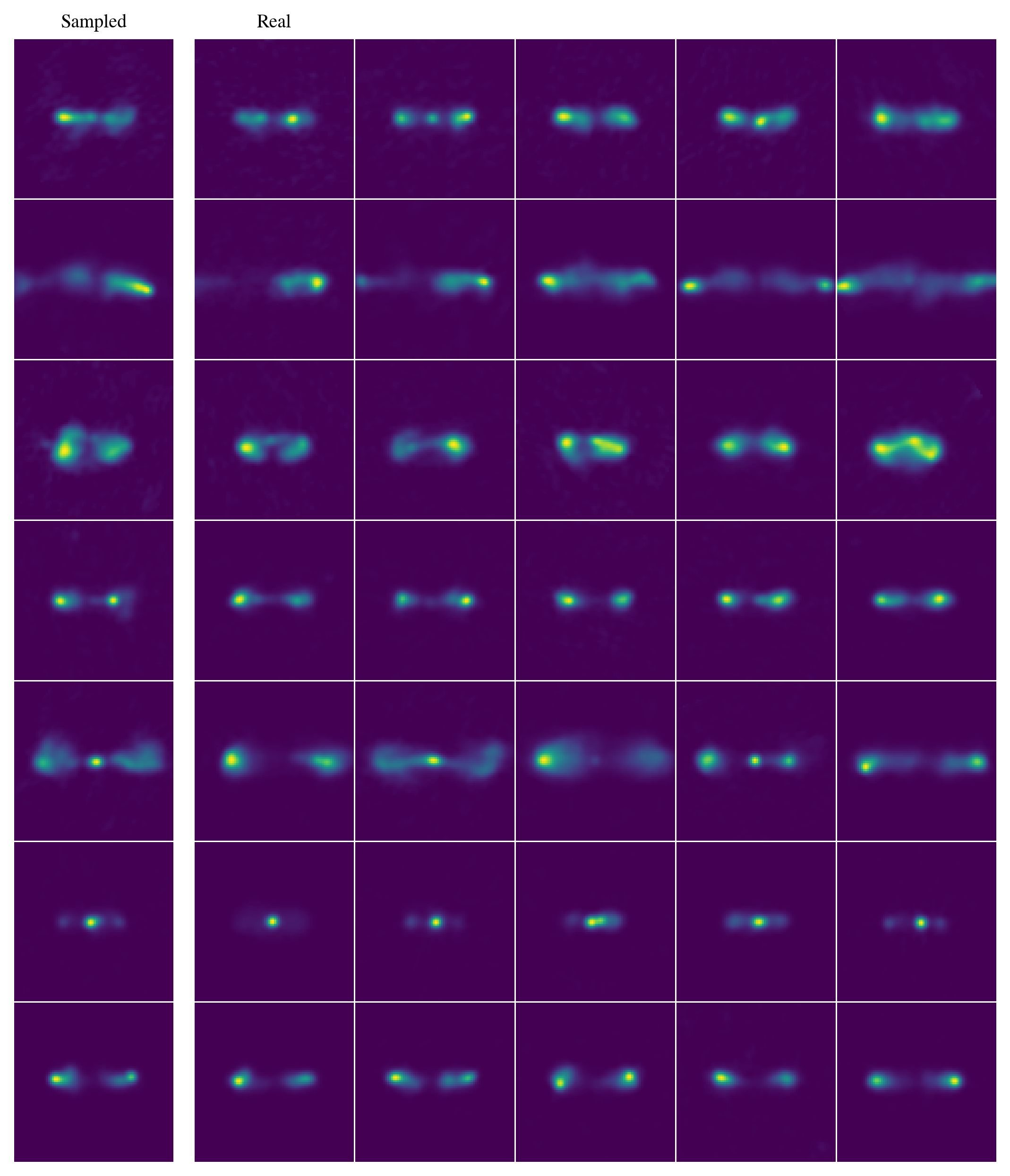}
    \caption{Sampled images (left panel) and five most similar images from the training dataset (right panel) as a first example.}
    \label{fig:app:nearest_neighbors_1}
\end{figure*}

\begin{figure*}
    \centering
    \includegraphics[width=17cm]{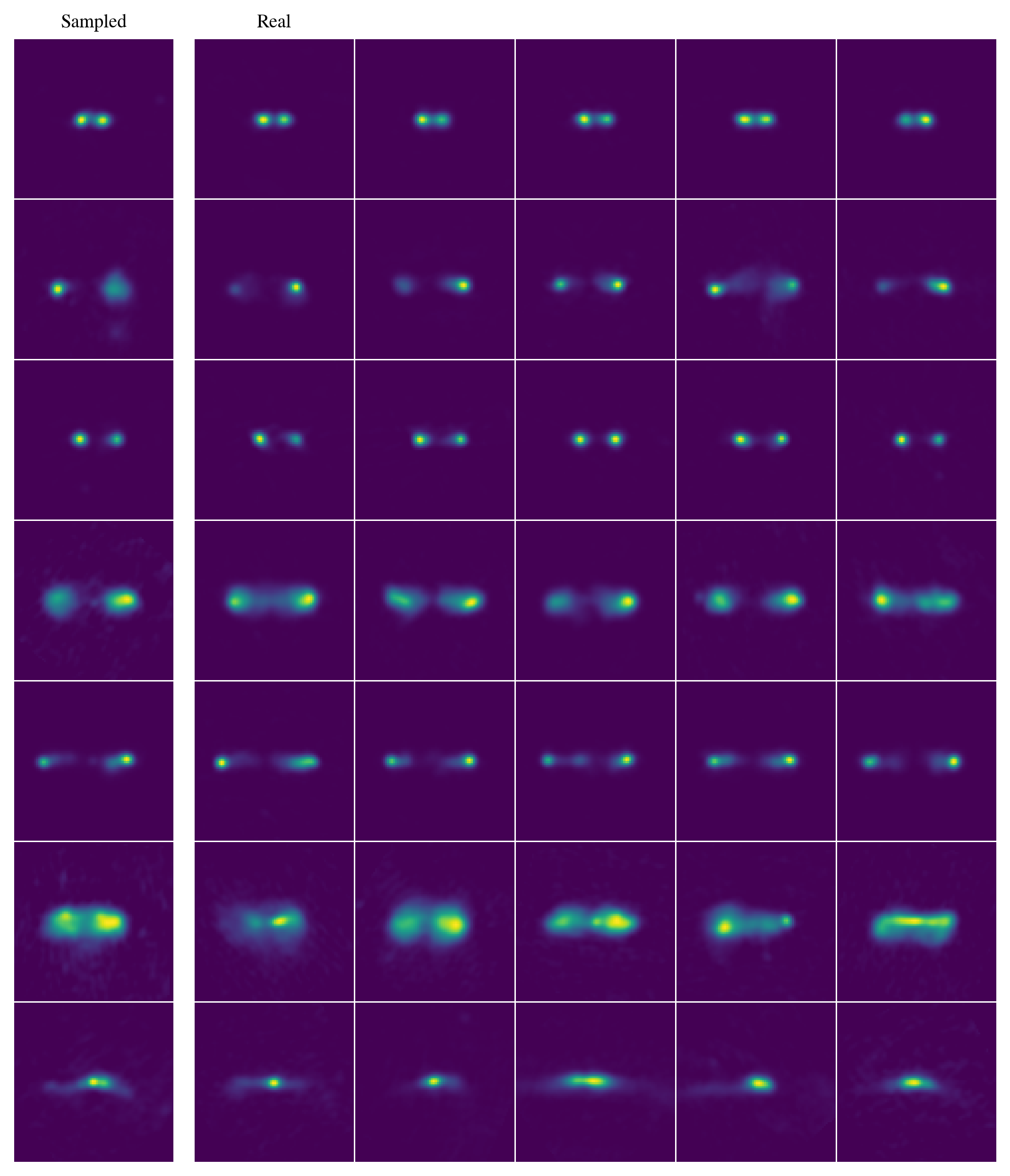}
    \caption{Sampled images (left panel) and five most similar images from the training dataset (right panel) as a  second example.}
    \label{fig:app:nearest_neighbors_2}
\end{figure*}

To illustrate the similarity between the generated samples and training data, we select a few generated samples and show the five most similar images from the training dataset for each. For this, all generated samples and training images are first aligned along their principal component, for which we consider pixels that exceed a threshold $\tau_\mathrm{PCA}$. This threshold is empirically set to
\begin{equation}
    \tau_\mathrm{PCA}(\mathbf{x}) = 0.1 + 0.5 \cdot \sigma_\mathbf{x} \label{eq:app:PCA_threshold}
,\end{equation}
for any given image $\mathbf{x}$, where $\sigma_\mathbf{x}$ is the standard deviation of the image, determined in the same way as for the $\mathit{S/N}_\sigma$ (see Sect. \ref{sec:sub:LOFAR_Data}). We subsequently run a principal component analysis (PCA) \citep{Pearson1901} using the selected pixels, and rotate the images such that the resulting principal component is aligned horizontally.\\
Image similarity is quantified with the structural similarity index measure ($\operatorname{SSIM}$) \citep{SSIM}, which for two images $\mathbf{x}, \mathbf{y}$ is calculated as 

\begin{equation}
    \operatorname{SSIM}(\mathbf{\mathbf{x}}, \mathbf{y})=\frac{\left(2 \mu_\mathbf{x} \mu_\mathbf{y}+c_1\right)\left(2 \sigma_{\mathbf{x} \mathbf{y}}+c_2\right)}{\left(\mu_\mathbf{x}^2+\mu_\mathbf{y}^2+c_1\right)\left(\sigma_\mathbf{x}^2+\sigma_\mathbf{y}^2+c_2\right)}, \label{eq:app:SSIM}
\end{equation}
where $\mu_\mathbf{i}$ and $\sigma_\mathbf{i}$ are the pixel mean and variance of image $\mathbf{i}$, and $\sigma_\mathbf{xy}$ is the covariance between the two images. The constants $c_1$ and $c_2$ are introduced to stabilize the division for small denominators and have values $c_1=\num{1e-4}$ and $c_2=\num{9e-4}$. The examples with highest-similarity matches are shown in Figs. \ref{fig:app:nearest_neighbors_1} and \ref{fig:app:nearest_neighbors_2}.

\end{appendix}

\end{document}